\documentclass[fleqn,usenatbib,useAMS]{mnras}

\usepackage{gensymb}
\usepackage{makecell}
\usepackage{graphicx}	
\usepackage{multicol}        
\usepackage{bm}		
\usepackage{pdflscape}	

\usepackage[T1]{fontenc}
\usepackage{ae,aecompl}

\usepackage{newtxtext,newtxmath}

\title{The merger fraction of post-starburst galaxies in UNIONS}

\author[Wilkinson et al.]{Scott Wilkinson$^{1}$\thanks{Contact e-mail: \href{mailto:swilkinson@uvic.ca}{swilkinson@uvic.ca}}, Sara L. Ellison$^{1}$, Connor Bottrell$^{2}$, Robert W. Bickley$^{1}$, Stephen Gwyn$^{3}$, \newauthor Jean-Charles Cuillandre$^{4}$, Vivienne Wild$^5$
\\
$^{1}$Department of Physics and Astronomy, University of Victoria, Victoria, British Columbia V8P 1A1, Canada
\\
$^{2}$Kavli IPMU (WPI), UTIAS, The University of Tokyo, Kashiwa, Chiba 277-8583, Japan
\\
$^3$Canadian Astronomy Data Centre, Herzberg Astronomy and Astrophysics, 5071 West Saanich Road, Victoria, BC, V9E 2E7, Canada
\\
$^4$AIM, CEA, CNRS, Université Paris-Saclay, Université de Paris, F-91191 Gif-sur-Yvette, France
\\
$^5$School of Physics and Astronomy, University of St Andrews, North Haugh, St Andrews KY16 9SS, UK}
\date{July 8$^\text{th}$, 2022}

\pubyear{2022}

\begin{document}
\label{firstpage}
\pagerange{\pageref{firstpage}--\pageref{lastpage}}
\maketitle

\begin{abstract}
    
    Post-starburst (PSB) galaxies are defined as having experienced a recent burst of star formation, followed by a prompt truncation in further activity. Identifying the mechanism(s) causing a galaxy to experience a post-starburst phase therefore provides integral insight into the causes of rapid quenching. Galaxy mergers have long been proposed as a possible post-starburst trigger. Effectively testing this hypothesis requires a large spectroscopic galaxy survey to identify the rare PSBs as well as high quality imaging and robust morphology metrics to identify mergers. We bring together these critical elements by selecting PSBs from the overlap of the Sloan Digital Sky Survey and the Canada-France Imaging Survey and applying a suite of classification methods: non-parametric morphology metrics such as asymmetry and Gini-M$_{20}$, a convolutional neural network trained to identify post-merger galaxies, and visual classification. This work is therefore the largest and most comprehensive assessment of the merger fraction of PSBs to date. We find that the merger fraction of PSBs ranges from 19\% to 42\% depending on the merger identification method and details of the PSB sample selection. These merger fractions represent an excess of 3-46$\times$ relative to non-PSB control samples. Our results demonstrate that mergers play a significant role in generating PSBs, but that other mechanisms are also required. However, applying our merger identification metrics to known post-mergers in the IllustrisTNG simulation shows that $\thicksim$70\% of recent post-mergers ($\lesssim $200 Myr) would not be detected. Thus, we cannot exclude the possibility that nearly all post-starburst galaxies have undergone a merger in their recent past.

\end{abstract}


\section{Introduction}
\label{intro}

Galaxies broadly fall into two categories, star-forming and quiescent, but can be further characterized by their morphology, colour, gas contents, and kinematics \citep{Strateva_2001, K03, Baldry04, Driver06, Wuyts2011, Bell12}. Cosmological simulations and observations agree that over billions of years, actively star-forming spiral galaxies are evolving into quiescent ellipticals, but the mechanisms causing this transition are complex and poorly understood \citep[e.g.,][]{BT87,LC93,SomDave15}. 

Identifying a population of galaxies that are actively transitioning from star-forming to quiescence allows us to study the mechanisms driving the quenching process. Post-starburst galaxies (PSBs)\footnote{Originally known as "E+A" or "K+A" galaxies due to having spectra typical for red ("K"), quiescent elliptical ("E") galaxies but with the unusual superposition of the spectral features of A-type stars ("+A").} offer such an opportunity. PSBs are characterized by spectra which indicate the presence of recently formed stars in a period of elevated star formation, yet no ongoing star formation \citep{DG83, CandS87}. The juxtaposition between recently formed stars and a lack of ongoing star formation indicates a rapid truncation in star formation. Additionally, their diverse morphologies \citep{Yang08,Meus2017,Pawlik18} and kinematic structure \citep{Pracy09, Pracy2013, Chen19} indicate that the rapid change in the star-formation of PSBs is coincident with a rapid change in the global properties of the galaxy. Thus, PSBs are viewed as an integral probe of rapid galaxy evolution.

 Cosmological simulations \citep[e.g.][]{Davis19, Pawlik19} show that a variety of mechanisms can cause galaxies to rapidly quench and become PSBs including ram pressure stripping and outbursts from active galactic nuclei (AGN). However, both major and minor mergers have been shown to be the most frequent cause of PSBs in simulations \citep{Davis19}. Furthermore, hydrodynamical simulations of individual galaxy mergers show that mergers can induce a burst in star formation followed by rapid quenching into a post-starburst phase, eventually progressing to become quiescent ellipticals \citep{Wild09, Snyder2011, Zheng2020}. While not all mergers go through a post-starburst phase, mergers with higher mass ratios, higher initial gas contents and prograde orbital configurations have been shown to induce stronger and longer lasting PSB signatures \citep{Bekki05,Wild09,Snyder2011,Pawlik18,Zheng2020}. Indeed, observationally, it is the most asymmetric mergers that exhibit the highest PSB fractions \citep{Rowlands18, Ellison22}.
 
 Observationally, PSBs have been shown to be present in dense clusters and caused by ram pressure stripping \citep{Werle22}. However, PSBs at low redshift are found predominantly in the field rather than in clusters \citep{Zab96, Quintero2004, Blake04, Goto05, Hogg06, Wild09, Yan09, Rowlands15} indicating that environmental quenching mechanisms like ram pressure stripping are unlikely to be the dominant cause of PSBs. Some PSBs are observed to have AGN \citep{Goto06,Tremonti07, Wild07, Wild10, Yesuf_2014} and AGN have been found to be more prevalent in interacting pairs and recent post-mergers \citep{Ellison19}. Significant fractions of PSBs are found to have irregular and disturbed morphologies indicative of a recent merger or major gravitational interaction with another galaxy \citep{Zab96, Blake04, Goto05, Yang08, Pracy09, Alatalo2016, Pawlik16, Pawlik18, Meus2017, Saz21}. Quantitatively, the merger fraction of post-starbursts has been determined to be anywhere from 13\% \citep{Blake04} to 60\% \citep{Pracy09}. However, numerous factors will impact the the merger fraction of the PSB population, including the precise definition of a "PSB" and how a merger is identified.

The observed morphology of a galaxy merger is highly sensitive to the image quality, particularly the depth and resolution of the image \citep{Lotz04,Pawlik16,B19-2}. In some cases, galaxy merger features may persist for up to $\thicksim$1 Gyr, but regularly begin to settle and become fainter on the order of $\thicksim$200 Myr after coalescence \citep{Mihos95, Lotz08, Bottrell22}. Considering that PSBs are typically 0.5 to 1 Gyr removed from the onset of the burst \citep{Wild10, Wild20}, the information regarding a PSB's (non-)merger history may have already faded beyond detection. Specifically, \citet{Pawlik16} found that the observed merger fraction of young PSBs reduced from 43\% (<300 Myr after starburst) to 21\% (>300 Myr after starburst) as the PSBs age. Due to the rapidly fading morphological information, the reported merger fractions of PSB samples are likely lower limits of the true fraction of recent mergers and interactions. Deeper imaging enhances low surface brightness features such as fading tidal tails and higher spatial resolution imaging allows for internal disturbances to be resolved \citep{Saz21}. Thus, higher quality imaging should allow for more mergers to be detected, approaching the true value of recent mergers. For example, when \citet{Yang08} revisited the same galaxies from \citet{Zab96} with deeper and higher resolution imaging from the Hubble Space Telescope, the detection of merger features increased from 5/21 (24\%) to 11/20 (55\%).

The diversity of PSB merger fractions reported in the literature is demonstrated in Table \ref{MergerSummary}. The wide range of reported values is caused by a heterogeneous selection of PSBs across different studies, small samples of PSBs, and differing merger identification methods (including the quality of imaging used). By selecting PSBs using different techniques, these studies are identifying slightly different populations for which the merger fraction may be different \citep[see Fig. 14 in][]{Meus2017}. Additionally, PSB merger fractions estimated from small samples (most previous studies contain only a few tens of objects) are subject to statistical error which contributes to the spread in reported values. Finally, visual identification of mergers is a subjective process and depends strongly on the quality of the imaging. Both the biases of the different classifiers and variation in the quality of imaging can be addressed if the merger identification methods are also applied to a robust control sample. Unfortunately, direct comparison to a control sample has not been commonly implemented. The purpose of the work presented here is to address many of these shortcomings by using 1) a large sample of PSBs, 2) high quality imaging, 3) a robust control sample, and 4) a broad suite of morphology metrics and merger identification methods. To address the differences that may incur from different selection methods, we study two samples of PSBs with differing selection criteria in tandem.

In this work, we use the Canada France Imaging Survey (CFIS) which offers deep and high-resolution imaging and has a large overlap with the Sloan Digital Sky Survey (SDSS) which offers optical spectra of sufficient quality for PSB selection (Section \ref{data}). This combination allows for a detailed morphological study (Section \ref{methods}) with a large sample of PSBs (Section \ref{ss}) of which we take advantage to quantify a robust merger fraction of PSBs (Section \ref{Results}). Finally, in Section \ref{IQ}, we test how the improved image quality offered by CFIS (compared to SDSS) affects our results.

\renewcommand{\arraystretch}{2.75}

\begin{table*}
\label{MergerSummary}
    
    \begin{tabular}{|l|c|c|c|c|c|c|c|c||}
    \hline
     & PSB Selection & Sample Size & Imaging$^\text{a}$ & Merger Identification & Merger Fraction & Merger Excess\\
    \hline
    
    \citet{Zab96} & \makecell{$\langle \text{H} \rangle^\text{b} >$ 5.5 Å \\ EW([OII]) $>$ -2.5 Å} & 21 & 
    \makecell{STScI Digitized\\ Sky Survey \\ 1.7 arcsec/pixel} & Visual Inspection & 24\% & No Controls\\

    \hline
    
    \citet{Blake04} & \makecell{$\langle \text{H} \rangle >$ 5.5 Å \\ EW([OII]) $>$ -2.5 Å} & 56 & \makecell{Supercosmos \\Sky Survey \\ 0.67 arcsec/pixel} & Visual Inspection & 13\% & No Controls \\
    
    \hline
    
    \citet{Blake04} & \makecell{EW(H$\delta$) > 5.5 Å \\ EW([OII]) $>$ -2.5 Å} & 71 & \makecell{Supercosmos \\Sky Survey \\ 0.67 arcsec/pixel} & Visual Inspection & 6\% & No Controls\\
    
    \hline
    
    \citet{Goto05} & \makecell{EW(H$\delta$) > 5.0 Å \\ EW([OII]) $>$ -2.5 Å \\ EW(H$\alpha$) $>$ -3.0Å} & 24$^\text{c}$ & \makecell{SDSS \\ 0.396 arcsec/pixel \\ } & Visual Inspection & 29\% &  No Controls\\
    \hline
    
    \citet{Yang08} & \citet{Zab96} & 20 & \makecell{HST \\ 0.128 arcsec/pixel \\ } & Visual Inspection & 55\% &  No Controls\\
    \hline

    \citet{Pracy09} & \citet{Zab96} & 10 & \makecell{Gemini \\ 0.145 arcsec/pixel} & Visual Inspection & 60\% & No Controls\\
    \hline
    
    \citet{Alatalo2016} & \makecell{EW(H$\delta$) > 5.0 Å \\ Shocked emission \\ CO Detection} & 52 & \makecell{SDSS \\ 0.396 arcsec/pixel \\ } & Visual Inspection & 37-46\% &  No Controls\\
    \hline

    \citet{Pawlik18} & \makecell{PC2>0 \\ Quiescent} & 29 & \makecell{SDSS \\ 0.396 arcsec/pixel \\ } & $A_S \geq 0.4$ & 14\% &  2.5$^\text{d}$\\
    
    \hline
    
    \citet{Pawlik18} & \makecell{PC2>0 \\ Quiescent} & 41 & \makecell{SDSS \\ 0.396 arcsec/pixel \\ } & Visual Inspection & 20\% &  2.9\\
    
    \hline
    
    \citet{Pawlik18} & \makecell{PC2>0 \\ Emission} & 49 & \makecell{SDSS \\ 0.396 arcsec/pixel \\ } & $A_S \geq 0.4$ & 10\% &  1.8\\
    
    \hline
    
    \citet{Pawlik18} & \makecell{PC2>0 \\ Emission} & 67 & \makecell{SDSS \\ 0.396 arcsec/pixel \\ } & Visual Inspection & 24\% &  3.6\\
    
    \hline
    
    \citet{Saz21} & \citet{Alatalo2016} & 26 & \makecell{HST \\ 0.128 arcsec/pixel \\ } & Internal Disturbance & 30\% &  --$^\text{e}$ \\
    
    \hline
    
    \citet{Saz21} & \citet{Alatalo2016} & 26 & \makecell{HST \\ 0.128 arcsec/pixel \\ } & $A_S \geq 0.4$ & 46\%$^\text{f}$ &  --\\
    
    \hline
    \hline
    
    This work. & \citet{Goto05} & 157 & \makecell{CFIS \\ 0.187 arcsec/pixel \\ } & $A_S \geq 0.4$ & 26\% & 2.9\\
    
    \hline
    
    This work. & \citet{Goto05} & 157 & \makecell{CFIS \\ 0.187 arcsec/pixel \\ } & \makecell{CNN \\ (p $>$ 0.5) \\ } & 30\% & 16\\
    
    \hline
    
    This work. & \citet{Goto05} & 157 & \makecell{CFIS \\ 0.187 arcsec/pixel \\ } & Visual Inspection & 42\% & 9.9\\
    
    \hline
    This work. & \makecell{PC2 $>$ 0 \\ PC1 $<$ -1.5} & 533 & \makecell{CFIS \\ 0.187 arcsec/pixel \\ } & $A_S \geq 0.4$ & 19\% & 2.6\\
    
    \hline
    
    This work. & \makecell{PC2 $>$ 0 \\ PC1 $<$ -1.5} & 533 & \makecell{CFIS \\ 0.187 arcsec/pixel \\ } & \makecell{CNN \\ (p $>$ 0.5)} & 16\% & 8\\
    
    \hline
    
    This work. & \makecell{PC2 $>$ 0 \\ PC1 $<$ -1.5} & 533 & \makecell{CFIS \\ 0.187 arcsec/pixel \\ } & Visual Inspection & 28\% & 9.6\\
    
    \hline

    \end{tabular}
    \caption{A summary of the methods and results from previous studies of the merger fraction of post-starburst galaxies in comparison to the work presented here. (a) Describing the quality of imaging used by each study in a brief table cell is not attainable. Here, we quote a descriptive title (for those familiar with the imaging programme or seeking more details) and the pixel scale resolution. While the pixel scale resolution is not more or less important than depth and atmospheric seeing, it is the only image quality metric used ubiquitously across all studies listed here and happens to correlate fairly well with the quality of imaging attained. (b) $\langle \text{H} \rangle$ is the average equivalent width of H$\beta$, H$\gamma$, and H$\delta$. (c) While \citet{Goto05} studies a sample of 266 "E+A" PSBs, only 24 were visually inspected. (d) As calculated from Table 4 of \citet{Pawlik18}. (e) \citet{Saz21} has a control sample but does not compute a merger excess. (f) As calculated using publicly available data from \citet{Saz21}. }
    \label{TAB}
    \end{table*}

\section{Data}
\label{data}

    \subsection{SDSS Imaging and Spectroscopy}
    
    \begin{figure}
        \centering
        \includegraphics[width=1\linewidth]{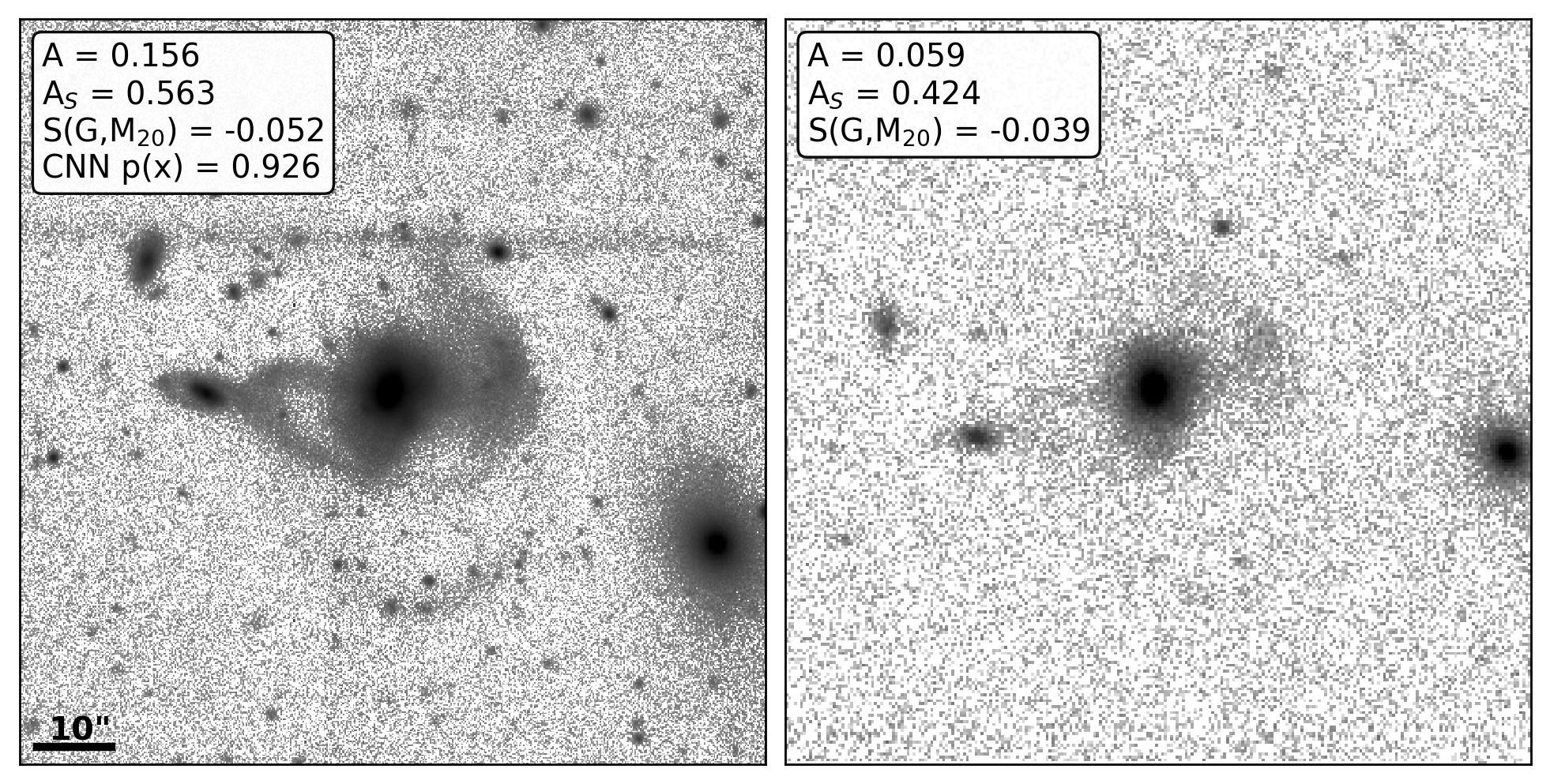}
        \caption{An example of a post-starburst galaxy with faint merger features as imaged by CFIS (left) and SDSS (right). The images are normalized and displayed on equivalent logarithmic scales. Relevant morphology parameters (see Section \ref{methods}) for merger classification are shown in the top left corners of each image. Although the merger features are much more evident in the CFIS image, based on the non-parametric morphology metrics both surveys would reach the same qualitative result: shape asymmetry identifies this as a merger while asymmetry and the Gini-M$_{20}$ merger statistic do not.}
        \label{cfisVsdss}
    \end{figure}
    
    The seventh data release of SDSS \citep[DR7;][]{SDSS-DR7} includes optical spectra and \emph{u}-, \emph{g}-, \emph{r}-, \emph{i}-, and \emph{z}-band images for over 900,000 galaxies in the northern sky\footnote{https://classic.sdss.org/dr7/}. In the work presented here, we make use of SDSS spectra and \emph{r}-band imaging. 
    
    SDSS spectra have a spectral resolution of R $\thicksim$ 800-2200 over the wavelength range 3800-9200 Å and are collected from optical fibres which have an on-sky aperture of three arcseconds. Thus, the spectra only include light from the central $\thicksim$500 pc for lower-redshift ($z \thicksim 0.025$) galaxies and from the central $\thicksim$4 kpc for higher-redshift ($z \thicksim 0.25$) galaxies in the survey. SDSS \emph{r}-band imaging has a pixel scale resolution of 0.396 arcsec/pixel, a median seeing of 1.4 arcsec, and a 1-$\sigma$ sky noise (i.e. depth) of 24.3 mag/arcsec$^2$.
    
    SDSS has been widely used to study galaxies at scale and, as a result, the value of the survey has grown through the production of ancillary data products. In particular, the  Max-Planck-Institute for Astrophysics–John Hopkins
    University (MPA-JHU) data catalogues\footnote{https://wwwmpa.mpa-garching.mpg.de/SDSS/DR7/} provide photometric stellar mass estimates following the methods of \citet{K03} and \citet{Salim07} as well as star-formation rates (SFR) based on \citet{B04}. The stellar masses from the catalogue are aperture-corrected total stellar masses and the SFRs are measured primarily from H$\alpha$ emission for star-forming galaxies and from calibration of the 4000Å break in cases where there is no H$\alpha$ emission. For most PSBs, the SFR is measured using the latter. Due to the rapidly changing SFR of PSBs and the relatively long timescale of 4000Å break sensitivity to SFR, the 4000Å break-calibrated SFRs are likely overestimating the instantaneous SFR in PSBs. SFRs are used to primarily select star-forming control galaxies and only to provide context to the PSBs and therefore any overestimation of PSB SFRs has no impact our analysis. We also use the emission line fluxes computed by \citet{K03a} which have been further corrected for internal galactic reddening. In total, there are 674,448 galaxies in SDSS DR7 with the ancillary data listed above. 
    
    \subsection{CFIS Imaging}
    \label{2.2}
    
    The Ultraviolet Near Infrared Optical Northern Survey (UNIONS) is a collaboration of wide field imaging surveys of the northern hemisphere. UNIONS consists of the Canada-France Imaging Survey (CFIS)\footnote{https://www.cadc-ccda.hia-iha.nrc-cnrc.gc.ca/en/community/unions/ 
    MegaPipe\_CFIS\_DR3.html}, conducted at the 3.6-meter Canada France Hawaii Telescope on Maunakea, members of the Pan-STARRS team, and the Wide Imaging with Subaru HyperSuprime-Cam of the Euclid Sky (WISHES) team. Together, UNIONS will assemble deep imaging of the Northern sky in the \emph{u}-, \emph{r}-, \emph{i}- and \emph{z}-bands. In the work presented here, we use the CFIS \emph{r}-band imaging only.
    
    The second CFIS data release (CFIS DR2) has \emph{r}-band images of over 373 million sources spread over 2,800 deg$^2$ of northern sky, packaged into 12,600 0.25 deg$^2$ tiles. CFIS imaging offers superior depth and resolution relative to that of SDSS. CFIS has a pixel scale resolution of 0.187 arcsec/pixel and in the \emph{r}-band it has a median seeing of 0.69 arcsec and a 1-$\sigma$ sky noise of 25.4 mag/arcsec$^2$. The superior depth of CFIS images allows for the observation of low surface brightness features that would be undetectable in SDSS imaging. To exemplify the difference CFIS imaging makes, we present both the CFIS (left) and SDSS (right) \emph{r}-band image of a post-starburst galaxy that happens to have faint merger features in Figure \ref{cfisVsdss}. 
    
    Galaxies in CFIS are matched to those in SDSS using their on-sky positions with a matching tolerance of three arcseconds. Of the 674,448 galaxies in SDSS DR7, there are 168,477 galaxies with coverage in CFIS DR2.
    
    \section{Methods}
    \label{methods}
    
    \subsection{\texttt{statmorph} Morphologies}
    \label{statmorph}
    
    Several previous works have produced morphology catalogues for SDSS galaxies \citep[e.g.,][]{Simard11, Mendel14}. Here, we extend these past efforts in two ways. First, we determine quantitative morphologies for a broad range of metrics that are developed for the identification of mergers. Second, we determine these metrics in a homogeneous and reproducible way using deep and high resolution CFIS imaging. 
    
    To meet these objectives, quantitative morphologies for each galaxy are computed using the Python package \texttt{statmorph} \citep{RG19}. \texttt{statmorph} takes the background-subtracted image of the galaxy, binary segmentation map identifying pixels belonging to the target galaxy, a point spread function (PSF), and a weightmap (see Section \ref{SMimp} for more details) and returns 43 non-parametric morphological statistics and the seven parametric quantities needed to construct a single 2-D Sérsic profile \citep{Sersic63}. Although all of the non-parametric morphology metrics are well established in the literature, for completeness we briefly review the definitions of the metrics relevant for identifying mergers used herein and provide references for readers desiring more detailed descriptions.
    
    \subsubsection{Asymmetry}
    
    Asymmetry \citep{Conselice00}, $A$, quantifies the azimuthal asymmetry of a galaxy's light profile. To extract this information, the image is rotated by 180$^\circ$, subtracted from the original image and normalized by the total flux. To account for contributions to the asymmetry from the background noise, the average asymmetry of the background is subtracted from the total:
    
    \begin{equation}
        A = \frac{\Sigma |I_0 - I_{180}|}{\Sigma |I_0|} - A_{\text{bgr}},
        \label{asym}
    \end{equation}
    
    \noindent where $I_0$ is the flux of a pixel in the original image, $I_{180}$ is the flux of the same pixel after the image has been rotated by 180$^\circ$ about the centre of the galaxy, determined by minimizing the value of A, and $A_\text{bgr}$ is the average asymmetry of the background. The sum is carried out over all pixels within 1.5 Petrosian radii of the galaxy’s centre.
    
    Low values of asymmetry indicate the galaxy is very azimuthally symmetric, a common feature of early-type galaxies with spheroidal morphologies \citep{Conselice03}. Spiral galaxies inherently have slightly elevated asymmetries due to naturally occurring asymmetric features like dust lanes and clumpy star formation \citep{Conselice03}. Higher asymmetry values are common amongst galaxies exhibiting strong merger and post-merger signatures; \citet{Conselice03} suggests galaxy mergers are those with $A > 0.35$. 
    
    \subsubsection{Shape Asymmetry}
    \label{As_methods}
    
    Shape asymmetry \citep{Pawlik16}, $A_S$, is defined in the same way as asymmetry (see Equation \ref{asym}) but instead of each pixel having an intensity, $I$, each pixel is given a binary value: 1 if the pixel belongs to the galaxy, 0 otherwise. The binary mask used to measure shape asymmetry is distinct from the binary segmentation map provided to \texttt{statmorph} as input and is generated internally by \texttt{statmorph} following the method described in \citet{Pawlik16}. The point around which the image is rotated is the same as the point in the asymmetry measurement which minimizes the light-weighted asymmetry. 
    
    By calculating the asymmetry of the binary mask rather than the flux of the image itself, equal weight is given to pixels belonging to both the faint and bright regions of the galaxy. Comparatively, standard light-weighted asymmetry would weight the central region of the galaxy in Figure \ref{cfisVsdss} 200-500$\times$ as much as the tidal features around it. Hence, the shape asymmetry statistic is more sensitive to low surface brightness features, such as fading tidal streams. This is of particular importance since faint tidal features may be key to identifying late-stage mergers expected for our PSB sample. 
    
    \citet{Pawlik16} suggests that a threshold value of $A_S \geq 0.2$ can be used to reliably identify mergers in SDSS. It is important to note that the definition of shape asymmetry given in \citet{Pawlik16} has an additional factor of 1/2 compared to the definition of \citet{Conselice00} and, importantly, \citet{RG19}. When the factor of 1/2 is accounted for, the \citet{Pawlik16} recommended merger threshold becomes $A_S \geq$ 0.4. 
    
    \citet{Pawlik16} notes that shape asymmetry will depend on the limiting magnitude of the imaging. It is therefore necessary to re-assess an appropriate threshold for identifying mergers in CFIS imaging. The CFIS \emph{r}-band images of approximately 200 galaxies were visually inspected in order to identify merger features. We found that a threshold of $A_S \geq 0.4$ remained a good nominal threshold for the identification of merger features.
    
    \subsubsection{Gini-M$_{20}$}
    \label{GM20}
    
    The Gini coefficent \citep{Lotz04}, G, is defined as the mean of the absolute difference of the light curve from a uniform distribution where the variable $X$ describes the flux in each pixel and is ordered from lowest to highest flux:
    
    \begin{equation}
        \text{G} = \frac{1}{2\overline{X}n(n-1)}\sum_i^n(2i-n-1)|X_i|,
        \label{G}
    \end{equation}
    
    \noindent where $n$ is the number of pixels associated with the galaxy and $\overline{X}$ is the mean of flux of the pixels belonging to the galaxy.
    
    The Gini coefficient is independent of the location of the brightest pixel and tends towards unity if the light from the galaxy is concentrated in a small number of pixels. If all of the galaxy's light were to come from a single pixel, G $= 1$, and if the light is evenly distributed across every pixel in the galaxy, G $= 0$. For a galaxy with a recent burst of star formation in the central regions, we might expect an increase in the Gini coefficient. 
    
    Before defining M$_{20}$, we first introduce the second moment of total light distribution, M$_\text{tot}$. The second moment of light, related to the spacial variance of the light distribution, is the summation of the flux of each pixel $f_i$ multiplied by the squared distance from the centre:
    
    \begin{equation}
        \hspace{1mm}{\text{M}_\text{tot}} = \sum_i^n\text{M}_i = \sum_i^nf_i\left[ (x_i - x_c)^2 + (y_i - y_c)^2\right],
        \label{Mtot}
    \end{equation}
    
    \noindent where $x_c$ and $y_c$ are x- and y-coordinates of the centre, determined by selecting the x- and y-coordinates that minimize M$_\text{tot}$.
    
    M$_{20}$, then, is defined as the second moment of the brightest pixels that produce 20\% of the galaxy's light, normalized by the total second moment of the galaxy. If pixels are ordered from highest to lowest flux, M$_{20}$ is calculated as:
    
    \begin{equation}
        \text{M}_{20} = \text{log}_{10} \left( \frac{\sum_{i}\text{M}_i}{\text{M}_\text{tot}} \right), \text{while} \sum_if_i < 0.2f_\text{tot}.
        \label{M20}
    \end{equation}
    
    In tandem, the Gini coefficient and M$_{20}$ can be used to identify galaxies with large portions of the total light profile contained within a small number of spatially separated pixels. Thus, Gini and M$_{20}$ are particularly potent indicators of early-stage galaxy mergers. \citet{Lotz08} suggests that at low-redshift ($z < 1.2$), galaxy mergers have Gini and M$_{20}$ values such that G $ > -0.14\text{M}_{20}+0.33$. \citet{RG19} define the Gini-M$_{20}$ merger statistic, S(G, M$_{20}$), as the distance of a galaxy in the Gini-M$_{20}$ plane from the merger cutoff line defined by \citet{Lotz08}. By this definition, mergers are defined by having S(G, M$_{20}$) $> 0$.
    
    \subsubsection{\texttt{statmorph} Implementation}
    \label{SMimp}
    
    In order to compute quantitative morphologies for a galaxy with \texttt{statmorph}, several inputs are required:
    
    \begin{itemize}
    
        \item \emph{Image}: Each galaxy's image is a 100"$\times$100" cutout from a CFIS tile. The selected 100"$\times$100" field of view ensures that, for even the largest, low-redshift galaxies, the entire stellar light profile including all potential extraneous low surface brightness features are contained entirely within the image with ample surrounding sky for robust sky statistics to be determined.
        
        \item \emph{Weightmap}: The weightmap provided to \texttt{statmorph} is the map of inverse variance generated by the CFIS calibration pipeline for the same region of sky, converted to a standard deviation weightmap.
        
        \item \emph{Binary Segmentation Map}: The segmentation map is generated using \texttt{sep} \citep{Barbary2016}, the Python implementation of Source Extractor \citep{BA1996}. We tune the input parameters of \texttt{sep} such that identified sources must have a minimum of five contiguous pixels with values greater than 1.5$\times$ that of the fiducial background noise. Nearby sources, particularly foreground stars and close pairs, are deblended internally by Source Extractor \footnote{Implemented with input parameters \texttt{deblend\_nthresh }$ = 32$ and \texttt{deblend\_cont }$ = 0.05$.}. Sources identified that are not the primary target are included in the mask.
        
        \item \emph{Mask}: \texttt{statmorph} interprets the mask as pixels the user wishes to be ignored from the morphology calculations. The mask for each galaxy includes the pixels belonging to sources identified by Source Extractor that are not the primary target but still in the field of view, pixels with flux values lower than -50, and pixels with values of exactly 0. The latter two specifically mask out chip gaps and other survey artefacts.
        
        \item \emph{Point Spread Function}: The PSF is approximated as a 2-D Gaussian profile with a full-width at half-maximum (FWHM) specific to the atmospheric seeing experienced by each galaxy and in accordance with its region within the CFIS tile. The PSF is only used by \texttt{statmorph} when computing the best-fit Sérsic profile.
        
    \end{itemize}
    
    A unique galaxy cutout, weightmap, segmentation map and point spread function were generated for each galaxy in the overlap of CFIS DR2 and SDSS DR7 and passed to \texttt{statmorph}. 2.8\% of galaxies did not have successful \texttt{statmorph} fits, often due to incomplete imaging at the edge of CFIS coverage. An additional 8.5\% of the remaining galaxies raise a general morphology flag. Since the flag cannot be traced back to determine which morphology calculation caused the problem, we disregard all morphology parameters for any galaxy with such a flag. This leaves us with 149,923 galaxies with trustworthy CFIS morphology measurements and complementary SDSS spectroscopic data products. The 149,923 galaxies represent the parent sample from which we will later select our PSBs and controls.
    
\subsection{Convolutional Neural Network Post-Merger Classification}
\label{CNN}

    In addition to the non-parametric morphologies described in Section \ref{statmorph}, we make use of recent advances in machine learning that have developed artificial neural networks for the purposes identifying galaxy mergers \citep[e.g.,][]{Pearson19,Ferr20,Ferreira22, Bottrell22}. Accounting for observational effects in the training of the neural network has proved crucial to a network's performance \citep{Bottrell19CNNReal, C21}. Hence, we utilize a convolutional neural network (CNN) trained to identify mergers specifically in CFIS from \citet{Bickley21}. 
    
    The CNN is capable of detecting mergers in CFIS by learning and synthesizing morphological features indicative of recent mergers from the 100-1 run of the IllustrisTNG cosmological simulation \citep{TNG1,TNG2,TNG3,TNG4,TNG5}. The CNN learns from post-mergers identified using the merger trees created by \texttt{SUBLINK} \citep{RG15}, following the methodology of \citet{Hani2020}. The merging galaxies must have had their subhalos coalesce within the timeframe of one snapshot in the simulation \citep[$t_\text{post-merger} \lesssim 162$ Myr;][]{Patton2020}, occurred at a time where $z \leq 1$, and have stellar mass ratios $\mu \geq 0.1$. Furthermore, their merger remnants must have a total stellar mass between $10^{10} \text{ M}_\odot$ and $10^{12} \text{ M}_\odot$. The post-merger and non-post-merger training samples naturally include a range of early- and late-type morphologies and star-formation histories. Realistic survey effects are integrated into the training of the CNN using the \texttt{RealSim-CFIS} code\footnote{github.com/cbottrell/RealSimCFIS}, a version of \texttt{RealSim}\footnote{github.com/cbottrell/RealSim} \citep{Bottrell19CNNReal} with specifications unique to CFIS. For further details regarding the CNN's construction, training, and validation, we refer the reader to \citet{Bickley21}.
    
    The CNN is trained to classify galaxies as post-mergers or non-post-mergers using the simulated CFIS-realistic image of the galaxy only; no additional information such as photometry data or star-formation history of the galaxy is provided to the CNN. For each galaxy image, the network returns a floating point prediction, $p(x)$, between 0 and 1, quantifying the certainty in its post-merger prediction. Traditionally, $p(x) > 0.5$ is interpreted as a positive prediction (in this case, a positive post-merger prediction) and $p(x) < 0.5$ is interpreted as a negative result (in this case, a non-post-merger prediction). However, this threshold can be raised to enhance the purity of the post-merger sample \citep[e.g.,][]{Bickley21}. While some previous works have used CNNs to classify galaxies into their morphological types \citep[e.g., ][]{Dieleman15, HC15}, this CNN is designed to predict whether a given galaxy is a post-merger or non-post-merger only.
    
    When tested on synthetic images of IllustrisTNG galaxies with CFIS realism applied, the CNN achieves a success rate of post-merger identification of $\thicksim$89\% with no significant dependence of the recovery rate on galaxy properties such as mass ratio \citep{Bickley21}. Furthermore, approximately half of the false-positive post-mergers are pre-coalescence interacting galaxies that go on to merge within 500 Myr. For the purposes of connecting PSBs to merger and interaction origins, including pre-coalescence merging galaxies is not an issue.
    
    \citet{Bickley22} have applied the neural network to the 168,477 galaxies in the overlap of CFIS DR2 and SDSS DR7, generating floating point post-merger predictions for each galaxy. The authors of \citet{Bickley22} further distilled 2000 galaxies with floating point predictions $p\gtrsim0.75$ into a pure sample of 699 visually-confirmed post-mergers. Using the raw floating point predictions allows us to identify mergers with high completeness, whilst using the visually-confirmed sample of post-mergers allows us to identify mergers with high purity. Hence, both the $p(x)$ values and the visually confirmed sample are used in Section \ref{CNNRes} to compute a merger fraction of PSBs.

\subsection{Visual Classification}
\label{VCmeth}
    
    Due to the vast and ever-increasing amount of available astronomical data and the variability of human classification leading to a lack of reproducibility, the field is moving away from visual classification as a core analysis method. Yet, our two PSB samples (consisting of 157 and 533 galaxies, respectively, and the selection of which is described in the following section) are still small enough that visual classification of the entire sample is a reasonable endeavour. The identification of mergers from the sample by-eye can then be compared to the various automated results as well as provide a more consistent comparison to many previous works \citep[e.g.,][]{Zab96,Blake04,Goto05,Yang08}.
    
    For this analysis, each galaxy is classified as either (1) a single galaxy with a disturbed morphology indicative recent merger or interaction, (2) a galaxy with a clear ongoing interaction with a companion galaxy, or (3) possessing no clear merger features. Galaxies classified as non-mergers may include isolated galaxies, galaxies with a companion in the frame but with no obvious interaction features (possibly due to either projection effects or a pre-pericentric companion), as well as galaxies for which the correct classification is unclear. The visual classification of post-starbursts and controls (the selection of which is described in the following section) is done in a randomized order, so as to remove any subconscious bias. The contrast, scaling and orientation of the \emph{r}-band image cutout are free to be manipulated by the classifier (SW) but no additional information such as mass, colour or redshift is made available. 
    
\section{Post-Starburst Sample Selection}
\label{ss}

    Traditionally, "E+A" post-starburst galaxies are selected on the basis of strong Balmer absorption lines (such as H$\delta$ and H$\gamma$) and weak nebular emission lines (such as H$\alpha$ and [OII]) \citep[e.g.,][]{Blake04,Goto05, Goto07}. Strong Balmer absorption is indicative of significant light contributions from A- and F-type stars. A-type stars have a lifetime of $\thicksim$1 Gyr and so their presence implies recent star formation. On the other hand, H$\alpha$ and [OII] are emitted by regions of active star formation. Thus, their absence implies that, although star formation has occurred in the last $\thicksim$1 Gyr, there is no ongoing star formation at the time of observation. Together, these spectroscopic indicators describe a galaxy that had a recent burst in star formation that has subsequently ceased quite rapidly.

    However, the traditional selection of PSBs requires strict cuts in nebular emission lines that excludes galaxies with some ongoing star formation and galaxies with other ionization methods such as active galactic nuclei (AGN) and shocks. Moreover, galaxies that have recently started to quench their star formation may have some remnant star formation occurring and should be included. Furthermore, both AGN and shocked emission  have been shown to be present in significant numbers of PSBs \citep{Tremonti07, Wild07, Yesuf_2014,Alatalo16-SPOGsampledescription,Alatalo2016, Saz21}. Principle component analysis (PCA) methods have proven capable of selecting PSBs without strict cuts in nebular emission which allow for the inclusion of PSBs with AGN, shocks, and residual star formation \citep{Wild07, Pawlik16, Pawlik18}. In the work presented here, we assess the merger fraction of both traditionally selected (no emission lines) and PCA selected (emission lines permitted) PSBs.

    \subsection{PSBs from Traditional Selection Methods}
    
    In order to assemble a sample of PSBs using a more traditional selection method, we draw from a pre-existing catalogue\footnote{Accessed at http://www.phys.nthu.edu.tw/$\thicksim$tomo/cv/index.html} of "E+A" PSBs in SDSS DR7 generated by \citet{Goto07}. The catalogue uses the following criteria to classify 816 galaxies as PSBs:
    
    \begin{itemize}
        \item[] EW(H$\delta$) > 5.0Å
        \item[] EW([OII]) > -2.5Å\footnote{Convention dictates that emission lines have negative equivalent widths while absorption lines are positive.}
        \item[] EW(H$\alpha$) > -3.0Å
        \item[] \emph{r}-band S/N > 10
    \end{itemize}
    
    For the purposes of our morphology analysis, we further require that the PSBs drawn from this catalogue have mass and redshift measurements from SDSS and CFIS imaging with flag-free \texttt{statmorph} morphologies, of which there are 169. 
    
    In the analysis presented later in this paper, we will quantify the fraction of PSBs that are classified as mergers (using the metrics defined in Section \ref{methods}). In order to contextualize the PSB merger fraction, it is necessary to also assess the merger fraction in a non-PSB control sample. If PSBs are triggered by mergers, we expect to see an elevated merger fraction in the PSB sample compared with the control sample.
    
    \begin{figure}
        \centering
        \includegraphics[width=1\linewidth]{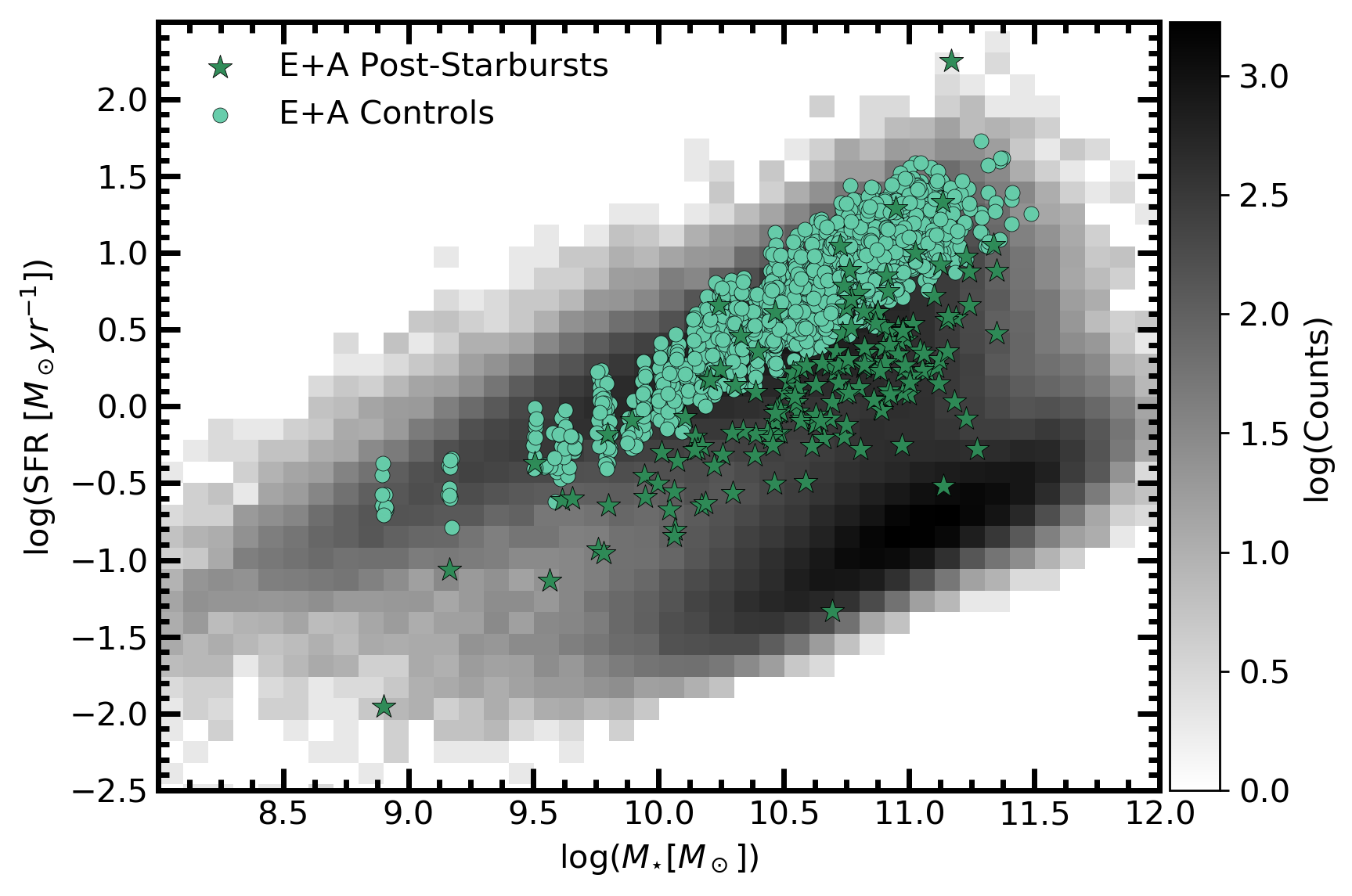}
        \caption{The location of the E+A PSBs (green stars) and their star-forming controls (light green circles) in the context of the star forming main sequence. The parent sample of $\thicksim$ 150,000 galaxies (with SDSS data products and flag-free CFIS morphologies) is shown as a 2-D histogram where the grey scale represents the number of galaxies in each bin.}
        \label{SFMS}
    \end{figure}

\begin{figure}
        \centering
        \includegraphics[width=1\linewidth]{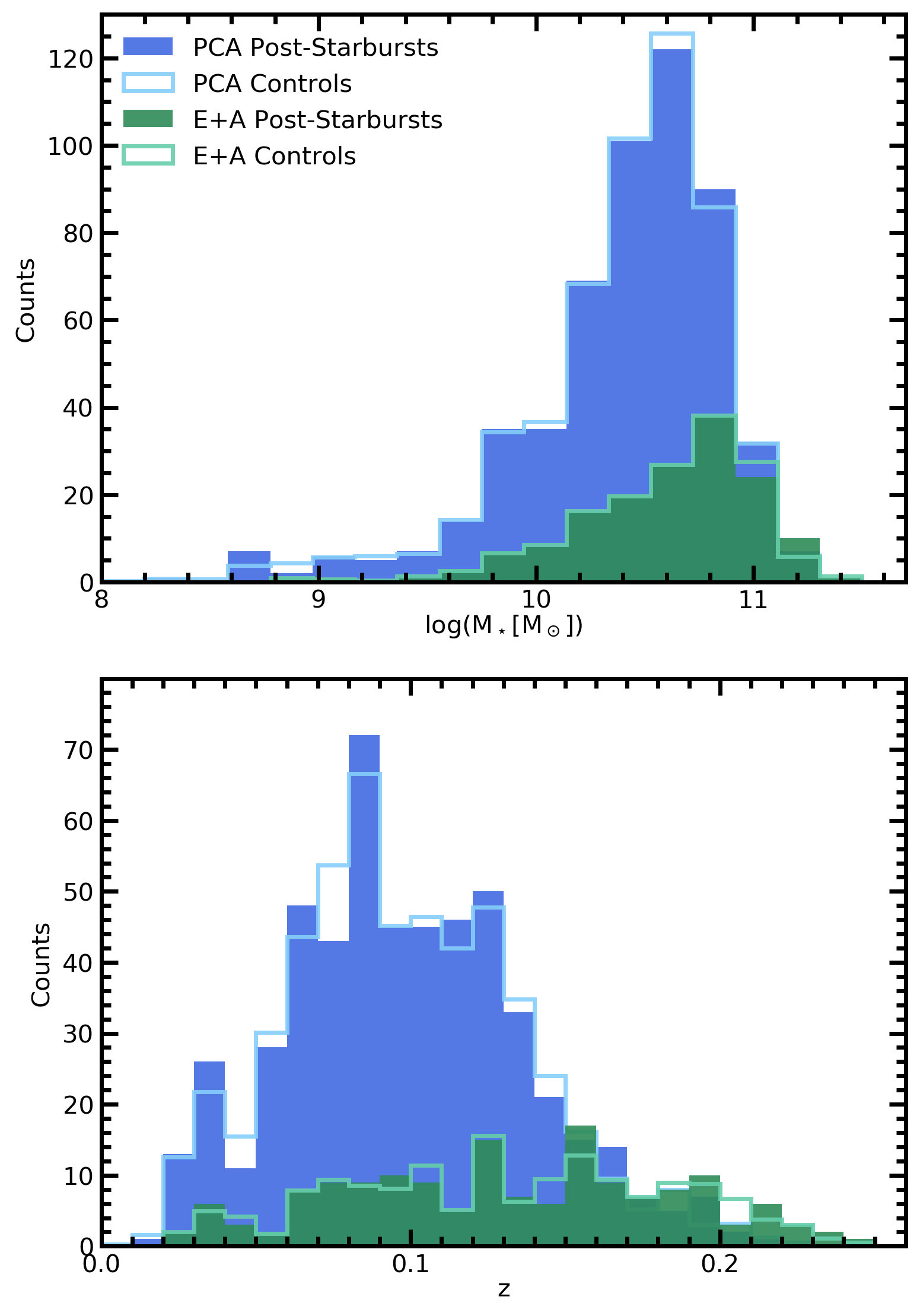}
        \caption{The stellar mass (top panel) and redshift (bottom panel) distributions of the PCA PSBs (blue), PCA star-forming controls (light blue lines), E+A PSBs (green), and E+A controls (light green lines). Note that since there are ten controls for each post-starburst, the histograms of the controls are weighted by a factor of $1/10$ so they can be compared directly to the post-starburst distributions.}
        \label{Mass-z}
    \end{figure}

    The control sample should represent the progenitors of the PSB sample. In this way, we can fairly compare the merger fraction of the pre-starburst event to the post-starburst remnant. Therefore, the pool from which we select star-forming controls is defined as being within 1$\sigma$ ($\thicksim$0.24 dex) of the SFR-M$_\star$ star-forming main sequence, excluding galaxies with emission lines indicative of AGN or LINERs as governed by the \citet{K03} parameterization. We anticipate that the number of identified mergers in this control sample will be quite low. Thus, it is imperative that the control sample has ample size for robust conclusions despite the difficultly of counting statistics of rare events. Therefore, each PSB is matched to ten star-forming galaxies of approximately equal stellar mass and redshift. This is done by iterating ten times through the PSB sample, each time selecting the closest match in log($M_\star$)-log($z$) space from the pool of controls. If in any of the ten iterations a PSB does not have a matched control within the matching tolerance of 0.2 dex in mass and 0.02 in redshift, that PSB is removed from our final sample and its previously matched controls are returned to the pool. There are 157 PSBs that successfully match ten controls which typically differ by much less than the maximum tolerance; the average absolute difference between the PSBs and their controls is $\Delta M_\star = 0.011$ dex and $\Delta z = 0.003$. The positions of these PSBs and their controls in relation to the star-forming main sequence in shown in Figure \ref{SFMS} and the resulting stellar mass and redshift distributions are shown in Figure \ref{Mass-z}. Henceforth, we will refer to the sample selected in this way as E+A PSBs.
    
    \subsection{PSBs from Principal Component Analysis}
    \label{2.1}
    
    PCA is a technique wherein a 1-dimensional spectra with $n$ data points is represented as a single point in $n$-dimensional space. A large sample of spectra such as SDSS DR7 becomes a cloud of points in $n$-dimensional space and orthogonal lines of greatest variance are considered eigenspectra onto which all other SDSS DR7 spectra are projected. The eigenspectra are numbered in order of the amount of variance each contributes to all spectra in SDSS. The projections onto the $1^\text{st}$ and $2^\text{nd}$ eigenspectra give values for principle components 1 and 2 (PC1 and PC2). The first two principle components carry enough information such that the evolutionary state of a galaxy can be roughly determined from only two components. PC1 captures very similar information as the 4000Å-break strength (D$_n$4000) and conveys information about the age of the galaxy's stellar population while PC2 is approximately equivalent to a measure of \emph{excess} Balmer absorption given the age of the stellar population and, as such, can be used to differentiate PSBs from the broader population. 
    
    To construct a more inclusive sample of PSBs than is afforded by traditional (no emission line) methods, we utilize the PCA catalogue\footnote{Accessed at http://www-star.st-and.ac.uk/~vw8/downloads/DR7PCA.html} of SDSS DR7 spectra from \citet{Wild07}. In order for the PCA to be reliable, the spectra require a g-band S/N $> 8$ \citep{Wild07,Pawlik18}. Hence, we only consider galaxies above that threshold. To select PSBs, we employ the same cut as in \citet{Pawlik18}:
    
    \begin{itemize}
        \item[] $PC2 - PC2_\text{err} > 0.0$.
    \end{itemize}
    
    Using only this cut, however, includes a significant amount of quiescent galaxies with high values of PC1 in our PSB sample. While these galaxies may have excess Balmer absorption given the age of their stellar population, the high stellar population age indicates that these galaxies have quenched their star formation long ago. This is not the population that we wish to study and so we make an additional cut:
    
    \begin{itemize}
        \item[] $PC1 + PC1_\text{err} < -1.5$.
    \end{itemize}
    
    \begin{figure}
        \centering
        \includegraphics[width=1\linewidth]{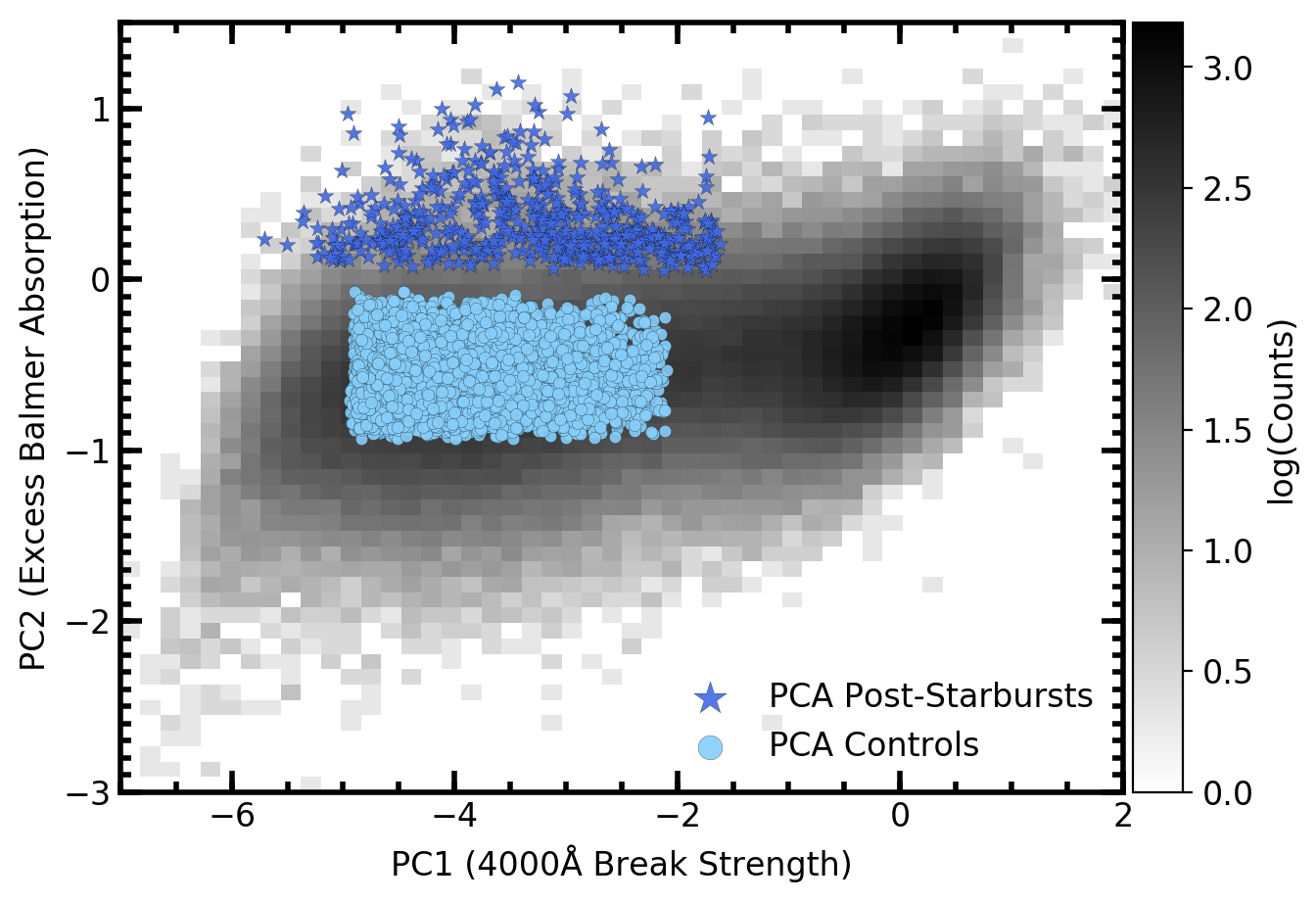}
        \caption{The location of the PCA PSBs (blue stars) and their star-forming controls (light blue circles) in the PC1-PC2 plane. The parent sample from which they are selected is shown as a 2-D histogram where the grey scale represents the number of galaxies in each bin. To the right of the figure, at high values of PC1, galaxies are typically quiescent ellipticals and in the bottom left corner, galaxies are typically recent starbursts \citep{Wild07}.}
        \label{PC1PC2}
    \end{figure}
    
    There is uncertainty in the existing literature about whether post-starbursts with high dust contents are truly post-starbursts or star-bursting galaxies with light from O- and B-type stars being preferentially obscured by dust \citep{Smail99,PogWu00, Miller2001, Goto04,Wild07,Yesuf_2014,Pawlik18}. \citet{Nielsen12} has shown that this is not a problem for the selection method used in \citet{Goto05} but \citet{Pawlik18} found that, based on SDSS imaging, the morphologies of PCA-selected "dusty PSBs" were consistent with their star-forming controls. Hence, we must recognize that the population of dusty PSBs studied by \citet{Pawlik18} is too ambiguous to be included in our sample and must be removed. 
    
    We quantify the dust content in a galaxy using its Balmer decrement. The Balmer decrement is the ratio between the strength of the redder H$\alpha$ emission line to the bluer H$\beta$. Since dust preferentially scatters shorter wavelengths, galaxies with Balmer decrements higher than the natural ratio of 2.86 have more dust \citep{Osterbrock2006}. The signal to noise of the Balmer decrement is defined as:
    
    \begin{equation}
        S/N = \frac{1}{\sqrt{(1/(S/N)_{H\alpha})^2+(1/(S/N)_{H\beta})^2}}.
    \end{equation}
    
    Following \citet{Pawlik18}, we require PSBs with nebular emission (which we define as having Balmer S/N > 3) to meet the following criteria: 
    
    \begin{itemize}
        \item[] $H\alpha/H\beta < 6.6$ if M$_\star$ $> 3\times10^{10}M_\odot$, or
        \item[] $H\alpha/H\beta < 5.2$ if M$_\star$ $< 3\times10^{10}M_\odot$.
    \end{itemize}

    As for the E+A PSBs, we must compile a control sample of star-forming non-PSBs for the PCA sample. Since the PCA PSBs are defined in PC1-PC2 space, we also draw controls from the PC1-PC2 plane. Following the method employed by \citet{Pawlik18} but with different bounds, the pool from which the PCA controls are drawn is restricted to the following regime of the PC1-PC2 plane:
    
    \begin{itemize}
        \item[] $PC1 - PC1_\text{err} > -5.0$
        \item[] $PC1 + PC1_\text{err} < -2.0$
        \item[] $PC2 - PC2_\text{err} > -1.0$
        \item[] $PC2 + PC2_\text{err} < 0.0$.
    \end{itemize}
    
    Each PCA-selected PSB is matched in stellar mass and redshift to ten star-forming controls using the same method outlined in the previous subsection. 533 PCA PSBs are successfully matched to ten controls. The average absolute difference between the PSBs and their controls is $\Delta M_\star = 0.017$ dex and $\Delta z = 0.003$. The redshift and stellar mass distribution of the matched samples is shown in Figure \ref{Mass-z} and their locations in the PC1-PC2 plane are shown in Figure \ref{PC1PC2}. Henceforth, we will refer to the sample of PSBs selected in this way as PCA PSBs.

\section{Results}
\label{Results}
     
    The primary objective of the work presented here is to assess the fraction of mergers in the PSB samples and to quantify whether this fraction is in excess of expectations given the matched control samples. The merger fraction of each sample is assessed with three approaches: quantitative morphologies computed with \texttt{statmorph} (Section \ref{QuantMorph}), CNN post-merger predictions (Section \ref{CNNRes}), and visual classification (Section \ref{VCRes}).
    
    To quantitatively compare PSBs to their controls, we use the merger fraction, $f$, and excess disturbance frequency, $Q$. For a given continuous morphology metric, $X$, that we wish to use as a merger indicator, a threshold, $T$, must be set beyond which galaxies are considered to be a merger. Hence, the merger fraction, $f$, and excess disturbance, $Q$, are defined as:
    
    \begin{equation}
        f_{X} = \frac{N(X>T)}{N},
    \end{equation}
    and
    \begin{equation}
        Q_{X} = \frac{f_{X,\text{ PSBs}}}{f_{X,\text{ Controls}}}.
    \end{equation}
    
    \subsection{Quantitative Morphologies}
    \label{QuantMorph}
    
\begin{figure*}
        \centering
        \includegraphics[width=\linewidth]{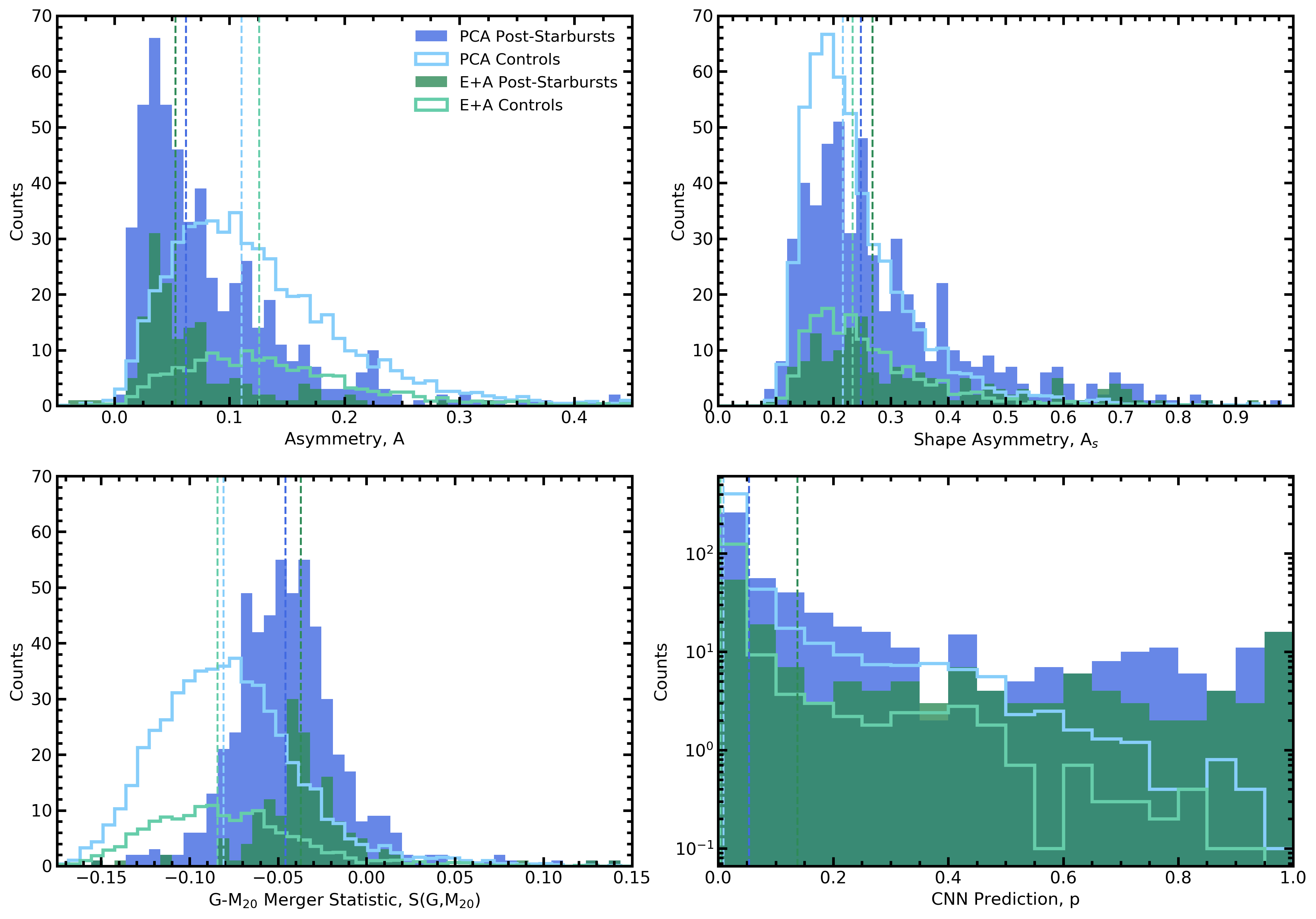}
        \caption{The distributions of four morphology parameters relevant to identifying mergers for the PCA PSBs (blue), PCA-controls (light blue), traditionally selected PSBs (green) and their controls (light green). In each figure, the medians of each population are plotted as vertical dashed lines and the values of those medians can be found in Table \ref{TAB}. Note that since there are ten controls for each post-starburst, the histograms of the controls are weighted by a factor of $1/10$ so they can be compared directly to the post-starburst distributions.}
        \label{dists}
    \end{figure*}

    In Figure \ref{dists}, we present a summary of the \texttt{statmorph}-derived morphologies of the PSB and control samples for metrics commonly used to identify mergers. In each panel, the distribution of the PCA PSBs and the E+A PSBs are shown with blue and green filled histograms, respectively. The distribution of their controls are shown as unfilled histograms of the corresponding lighter colours. Since the controls outnumber the PSBs by a factor of ten, the counts of the controls are weighted by a factor of 1/10. Presenting the data in this way preserves the relative size of the PCA and E+A PSB samples while also allowing for direct comparison between the distributions of the PSB samples and their controls. Vertical dashed lines indicate the medians of the distributions for which exact values can be found in Table \ref{TAB}. A more detailed description of each panel is described in the subsections that follow.
    
    \renewcommand{\arraystretch}{1.5}
    
    \begin{table*}
    \begin{tabular}{|l|c|c|c|c|c|c|}
    \hline
    
    Sample & A & A$_S$ & G & M$_{20}$ & S(G, M$_{20}$) & CNN $p$\\
    
    \hline
    
    PCA PSBs & 0.06$\substack{+0.08 \\ -0.03}$ & 0.25$\substack{+0.18 \\ -0.09}$ & 0.56$\substack{+0.04 \\ -0.04}$ & $-$1.94$\substack{+0.20 \\ -0.18}$ & $-$0.05$\substack{+0.03 \\ -0.02}$ & 0.05$\substack{+0.43 \\ -0.05}$  \\
    
    PCA Controls & 0.11$\substack{+0.08 \\ -0.06}$ & 0.22$\substack{+0.11 \\ -0.06}$ & 0.50$\substack{+0.05 \\ -0.05}$ & $-$1.82$\substack{+0.17 \\ -0.18}$ & $-$0.08$\substack{+0.04 \\ -0.04}$ & 0.01$\substack{+0.09 \\ -0.01}$ \\
    
    E+A PSBs & 0.05$\substack{+0.06 \\ -0.02}$ & 0.27$\substack{+0.26 \\ -0.09}$ & 0.57$\substack{+0.03 \\ -0.03}$ & $-$1.94$\substack{+0.13 \\ -0.18}$ & $-$0.04$\substack{+0.02 \\ -0.02}$ & 0.14$\substack{+0.64 \\ -0.13}$ \\
    
    E+A Controls & 0.13$\substack{+0.09 \\ -0.06}$ & 0.23$\substack{+0.12 \\ -0.07}$ & 0.50$\substack{+0.05 \\ -0.05}$ & $-$1.80$\substack{+0.20 \\ -0.18}$ & $-$0.08$\substack{+0.04 \\ -0.04}$ & 0.00$\substack{+0.09 \\ -0.00}$  \\
    
    \hline
    \end{tabular}
    \caption{The median value of several morphology parameters for each of the samples derived from CFIS imaging. The lower and upper errors on each median are calculated using the 16$^\text{th}$ and 84$^\text{th}$ percentiles, respectively. }
    \label{TAB}
    \end{table*}
    
    \subsubsection{Gini-M$_{20}$ Merger Statistic}
    \label{GM20-Results}
    
    In the lower left panel of Figure \ref{dists}, we show histogram distributions of the Gini-M$_{20}$ merger statistic for the PSB samples (solid histograms) and their respective controls (open histograms). Looking first at the results from the PCA-selected samples, we find that the PCA PSBs have elevated values of the Gini-M$_{20}$ merger statistic relative to their star-forming controls. However, at the merger threshold of S(G, M$_{20})>0$, the merger fraction of the PCA PSBs is only $f_{\text{G-M}_{20}} = (7.5\pm1.1)$\%, at least a factor of 2 lower than expected based on previous studies (see Table \ref{MergerSummary}). Nonetheless, the PCA PSBs have a higher merger fraction than that of their controls which have a merger fraction of $f_{\text{G-M}_{20}} =$ ($4.2\pm0.3$)\%. This translates to an excess over the controls of $Q_{\text{G-M}_{20}} = 1.8\pm0.3$ for PCA PSBs. 
    
    The result is much the same for the E+A PSBs for which the default Gini-M$_{20}$ threshold gives $f_{\text{G-M}_{20}} = (7.0\pm2.0)\%$ with an excess over the controls of $Q = 1.4\pm0.4$. The excess of mergers in the E+A PSBs over their controls is consistent with one, in part due to the statistical uncertainty that comes with a smaller sample size. Therefore, at the default threshold, the Gini-M$_{20}$ merger statistic does not seem to identify a statistically significant surplus of mergers in the E+A sample of PSBs.
    
    In theory, it is possible to explore the potential merger excess in the PSB sample using a more strict Gini-M$_{20}$ cut (i.e. larger values of S(G, M$_{20}$)). Indeed, with the other statistics ($A$, $A_S$, and the CNN prediction) we will perform such a test. However, the number of galaxies in any of the PSB or control samples with S(G, M$_{20})>0$ is quite small and so no statistically robust conclusions can be drawn. Furthermore, galaxies contaminated by a foreground star near its nucleus have highly concentrated and spatially separated light distributions that artificially enhance S(G, M$_{20}$), sometimes causing false-positive merger identifications. These cases are difficult to deblend or flag as doing so would also remove double-nuclei pre-coalescence merging galaxies and galaxies with clumpy star formation from our morphology catalogue. Foreground star contamination affects each population equally which is, in part, driving the excess to unity.
    
    From hydrodynamical simulations of individual galaxy mergers, \citet{Lotz08} find the Gini-M$_{20}$ merger indicator fades more rapidly than other morphology metrics. Thus, S(G, M$_{20}$) seems to identify ongoing mergers and interactions more effectively than late-stage post-mergers (this is discussed further in Section \ref{Efficacy}). For our PSBs, which are expected to be several 100 Myr removed from the starburst event,  S(G, M$_{20}$) does not seem to be a particularly salient merger indicator.
    
    \subsubsection{Asymmetry}
    \label{Asymmetry}
    
    Significant fractions of galaxies with moderate asymmetries in the control samples is to be expected; both late-type disks \citep[A  = 0.15 ± 0.06;][]{Conselice03} and edge-on disks \citep[A  = 0.17 ± 0.11;][]{Conselice03} will contribute a number of moderately high asymmetry measurements due to clumpy star formation and the edge-on view of dust lanes in disks. However, the asymmetry distributions presented in the top left panel of Figure \ref{dists} show that both samples of PSBs tend to have lower asymmetry measurements than their star-forming controls.  So, relative to their star-forming controls, PSBs actually have \emph{suppressed} asymmetries, and even more dramatically so in the case of E+A PSBs.
    
    Adopting a threshold of $A > 0.35$ to discriminate mergers from the sample gives a merger fraction of $f_A =$ ($1.9\pm0.6$)\% for the PCA PSBs, translating to an excess over their controls of $Q_A = 0.9\pm0.3$. Thus, the number of identified mergers in the PCA PSBs is consistent with no change from their star-forming controls. The merger fraction of the E+A PSBs is even lower at $f_A =$ ($1.3\pm0.9$)\%. In fact, there are fewer mergers in the E+A PSB sample than in their star-forming controls with a excess below one of $Q_A = 0.4\pm0.3$.
    
\begin{figure}
        \centering
        \includegraphics[width=1\linewidth]{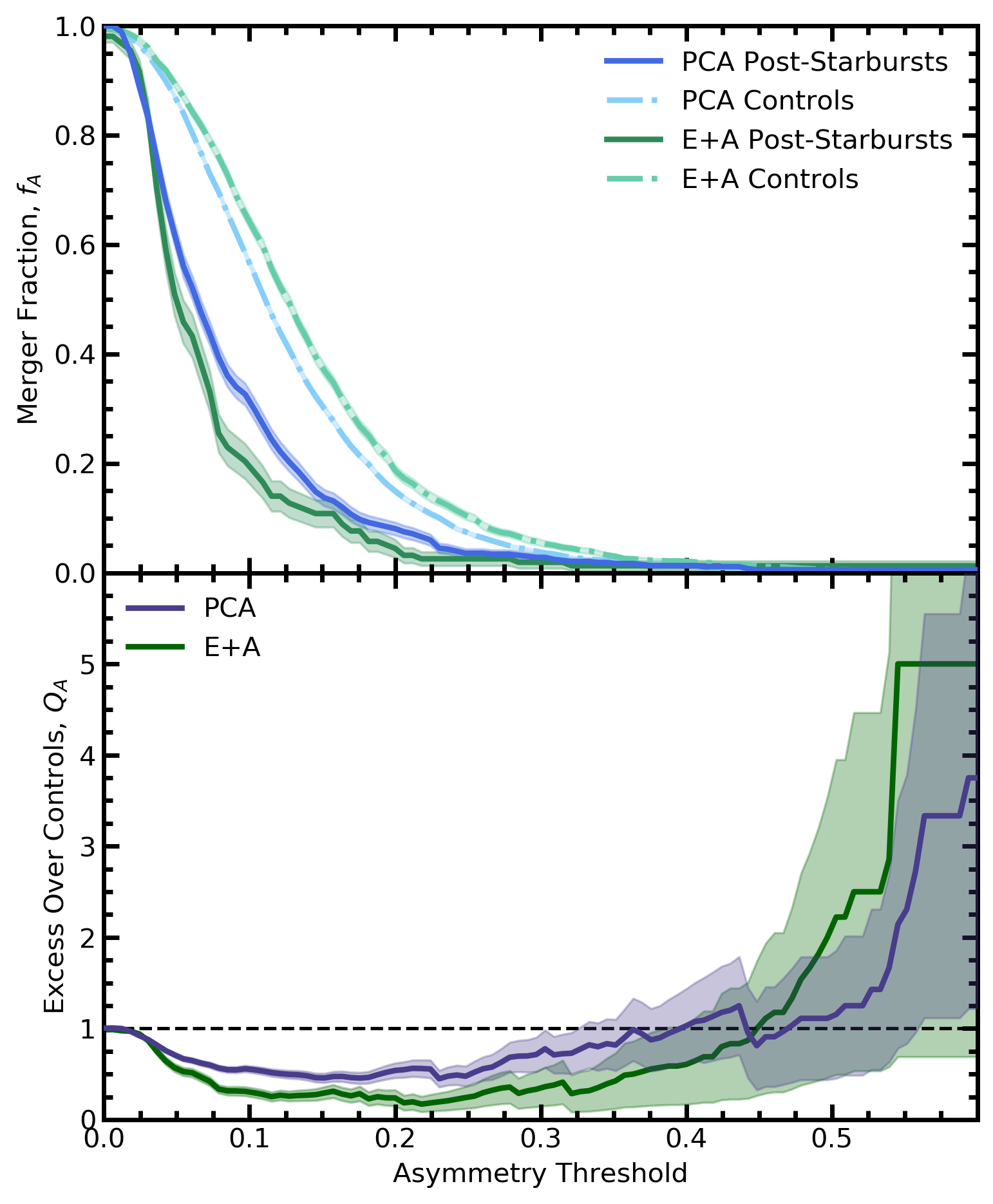}
        \caption{\emph{Top}: The merger fraction of PCA PSBs (blue line) and E+A PSBs (green line) compared to their star-forming controls (dotted lines) as a function of the threshold above which a galaxy would be considered a merger. For example, using a threshold of $A>0.1$ to identify mergers results in $\thicksim$20\% of E+A PSBs being classified as mergers. Since all galaxies typically have a positive asymmetry, 100\% of galaxies in all samples would be classified as mergers if a $A>0$ threshold were imposed.
        \emph{Bottom}: The excess fraction of mergers in each sample relative to their star-forming controls as a function of the asymmetry threshold used to identify mergers based on the asymmetry of their light profiles. The black dashed line represents the line of equality between the merger fraction of the PSBs and their controls.}
        \label{DRA}
    \end{figure}

    In Figure \ref{DRA}, we explore how our results are affected by the asymmetry threshold used to define a merger. In the top panel we present the merger fraction of the PCA and E+A PSBs as a function of the asymmetry threshold used to define a merger with solid blue and green curves, respectively. For comparison, we show on the same panel the merger fraction of the PCA controls (light blue) and E+A controls (light green). Shaded areas around the curves represent the binomial error of the merger fraction which are propagated into the lower panel. In the lower panel, we show the excess of mergers in the PSBs over their controls for the PCA PSBs (dark blue) and E+A PSBs (dark green). In this case, we find that there is an excess number of galaxies with very high asymmetries ($A\gtrsim0.5$) in both PSB samples but equal or fewer mergers in the PSB samples than their star-forming controls ($Q_A \lesssim 1$) for all other considered merger thresholds.
    
    The suppressed asymmetries of the PSB samples cause the merger fractions determined using asymmetry to be significantly lower than previous studies would suggest (see Table \ref{MergerSummary}) and lower than those identified using the Gini-M$_{20}$ merger statistic in Section \ref{GM20-Results}. However, there is an additional nuance not previously considered: PSBs selected on the basis of SDSS centrally-located optical fibres would, by definition, have had a recent burst in star formation in the central region of the galaxy. It is likely that the central burst in star formation would produce a very bright, azimuthally symmetric light profile at the centre of the galaxy which would increase the denominator of Equation \ref{asym} with negligible change to the numerator, systematically driving down the asymmetry calculation of each post-starburst galaxy (see Appendix \ref{central} for a more detailed discussion).
    
    If, due to the nature of the galaxies we are observing, our PSBs are expected to have bright, concentrated, azimuthally symmetric cores, ordinary asymmetry may not be the best metric to derive a merger fraction of PSBs as these bright, symmetric centres are weighted much more heavily than any faint, asymmetric tidal features. Perhaps shape asymmetry, which takes the asymmetry of the binary mask rather than the flux of each pixel in the image and thus distributes more statistical weight to faint tidal features, will be a more applicable metric for identifying mergers in the PSB populations.
    
    \subsubsection{Shape Asymmetry}
    \label{ShapeAsymmetry}
    
    \begin{figure}
        \centering
        \includegraphics[width=1\linewidth]{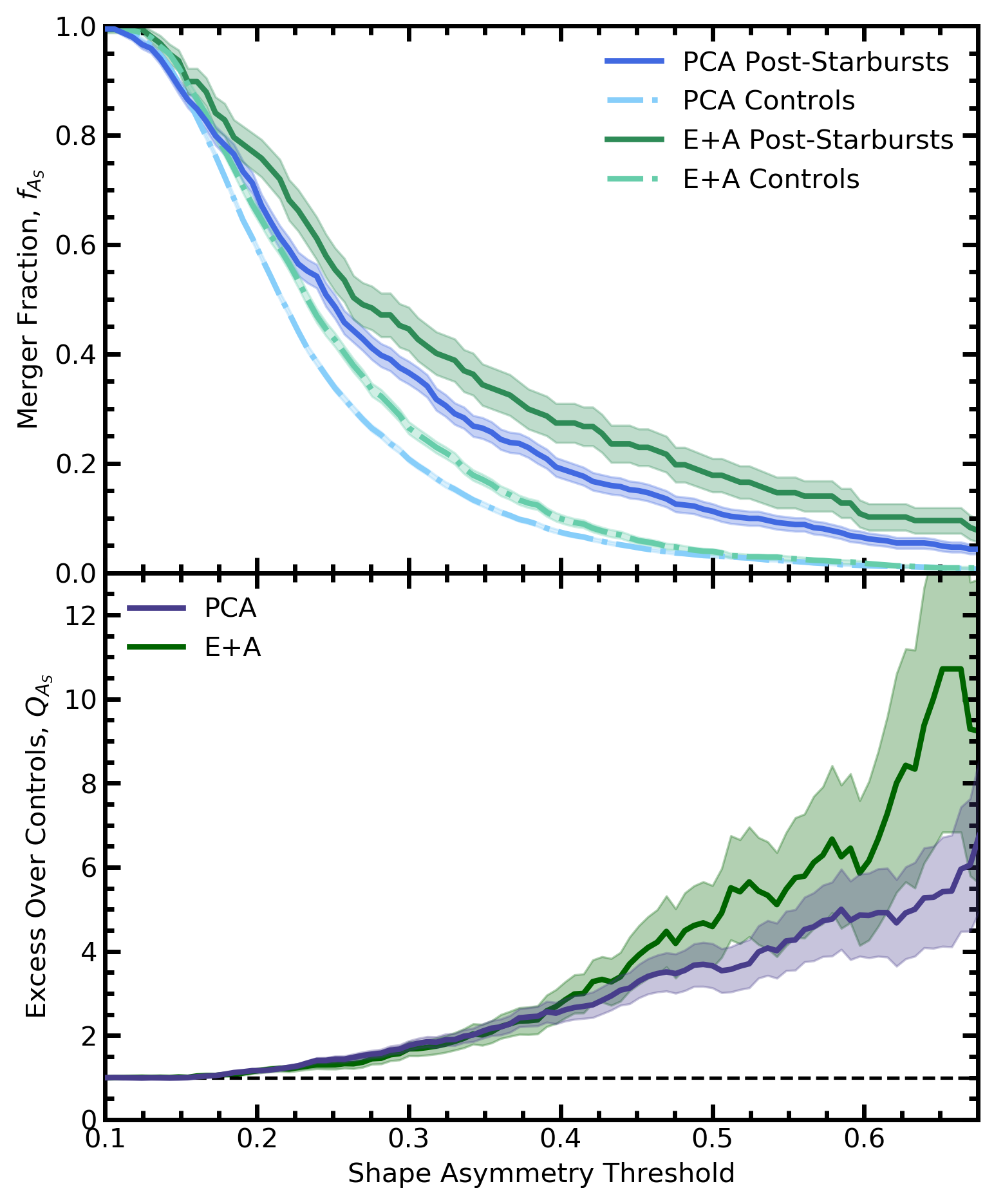}
        \caption{Same as Figure \ref{DRA} but considering shape asymmetry.}
        \label{AsDR}
    \end{figure}
    
    \citet{Pawlik16} suggest a threshold of $A_S > 0.2$ is appropriate for the identification of mergers from SDSS imaging. However, as noted in Section \ref{As_methods}, a factor of 2 difference in the definition of $A_S$ translates this to a threshold of 0.4 for the \texttt{statmorph} measurements. From a visual inspection of the CFIS images, we also confirm that this is an appropriate threshold for our sample. Adopting a threshold of $A_S > 0.4$, we find that the merger fraction for the PCA PSBs is $f_{A_S} = (19\pm2)\%$ with an excess over their controls of $Q_{A_S} = 2.6\pm0.3$. At the same threshold, the merger fraction of the E+A PSBs is $f_{A_S} = (26\pm3)\%$ which is significantly higher than that of the PCA PSBs, yet the excess over their controls is the same within statistical uncertainty at $Q_{A_S} = 2.8\pm0.4$.
    
    As was done previously for asymmetry to explore the effect of the threshold used to classify a galaxy as a merger, we calculate $f_{A_S}$ and $Q_{A_S}$ for a range of possible thresholds. The green and blue curves in the top panel of Figure \ref{AsDR} show that, regardless of our selected threshold, both samples of PSBs exhibit higher fractions of asymmetric low surface brightness tidal features than their controls (light blue and green dotted lines). Furthermore, the excess disturbance frequency is greater than one for both samples at all considered thresholds and the values of $Q_{A_S}$ for both samples steadily increase as the threshold is increased. This indicates that the fraction of very highly disturbed PSBs ($A_S > 0.6$) is in excess over the controls by a factor of $Q\thicksim 5$. Due to the sensitivity of shape asymmetry to faint tidal features expected for PSBs and the agreement of the shape asymmetry merger fraction with previous studies, shape asymmetry seems to be a more reliable automated merger detection method than using Gini-M$_{20}$ or asymmetry (see Section \ref{Efficacy} for a more detailed discussion).
    
    \subsection{CNN Post-Merger Prediction}
    \label{CNNRes}
    
    In the bottom right panel of Figure \ref{dists}, we show the distributions of the floating point post-merger predictions assigned to each sample by the CNN. Typically in machine learning binary classification problems, the default threshold above which the output is considered a positive prediction is $p>0.5$. The CNN finds that a majority of PSBs are assigned a floating point prediction less than 0.5 and therefore are not recent post-mergers. The distributions are so strongly clustered around the non-post-merger prediction (0) that we choose to display the distributions in log-scale so that differences between the samples at higher values of $p$ can be seen.
    
   At the default threshold of $p > 0.5$, the PCA PSBs have a merger fraction of $f_\text{CNN} = (16\pm2)\%$ and an excess over their controls of $Q_\text{CNN} = 8\pm1$. In contrast, at the default threshold of $p > 0.5$, the E+A PSBs have a merger fraction of $f_\text{CNN} = (30\pm4)\%$, in excess over their controls by a factor of $Q_\text{CNN} = 16\pm4$. These fractions indicate that post-merger galaxies are approximately twice as abundant in the E+A PSBs than in the PCA PSBs.
    
\begin{figure}
        \centering
        \includegraphics[width=1\linewidth]{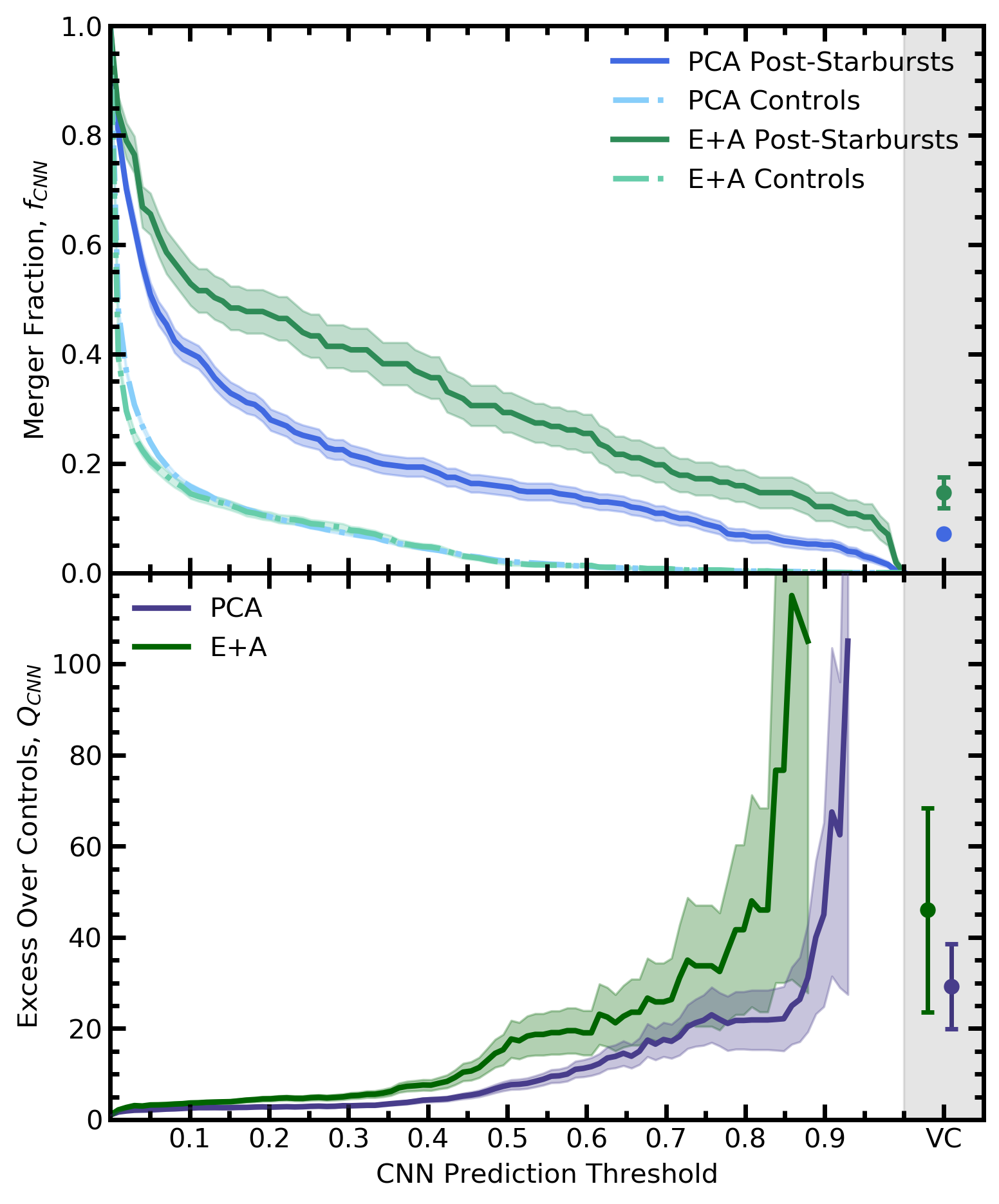}
        \caption{Same as Figures \ref{DRA} and \ref{AsDR} but considering the CNN post-merger predictions. Predictions closer to 1 are more likely to be mergers and predictions closer to 0 are more likely to be isolated galaxies. The shaded region of the plot is distinct and represents the number of galaxies with $p \gtrsim 0.75$ that were visually confirmed (VC) by \citet{Bickley22} to be post-merger galaxies.}
        \label{FCNN}
    \end{figure}

    Although a $p > 0.5$ is a standard decision threshold in machine learning and despite the high performance of the CNN, \citet{Bickley21} discuss how the application of this cut for a general dataset of galaxies is expected to lead to a merger sample that is only 6\% pure \citep[see also][]{Bottrell22}. The degree of impurity is likely to be less extreme for our sample which are expected to have an excess of mergers. We improve our assessment of the CNN merger fraction in two ways. First, in Figure \ref{FCNN} we explore how the value of $f_\text{CNN}$ would differ over a range of potential merger thresholds wherein higher values of $p$ indicate higher confidence in the post-merger prediction from the CNN leading to a more pure sample of post-mergers. Second, by counting only visually confirmed post-mergers with $p>0.75$ as identified post-mergers we distill a pure sample of post-mergers albeit one that is incomplete.

    The top panel of Figure \ref{FCNN} shows that regardless of the CNN prediction threshold used, the merger fraction of PSBs galaxies is significantly higher than that of their star-forming controls and even higher for the E+A PSBs. The merger fraction in both samples decreases gradually as the threshold increases. This signifies a gradual exchange of completeness for purity when progressively higher decision thresholds are used. Thus, using any choice of threshold above which galaxies will be considered mergers will be somewhat arbitrary.
    
    Using thresholds higher than $p > 0.5$ to classify galaxies as mergers increases the excess over the controls significantly. At very high thresholds ($p \gtrsim 0.85$) where the CNN has given galaxies a very high probability of being a post-merger, very few control galaxies are classified as post-mergers with this high confidence and as a result the excess of post-merger galaxies in both samples of PSBs is on the order of $Q\thicksim100$ ($\pm70$).
    
    The shaded area of Figure \ref{FCNN} indicates the number of visually confirmed mergers from \citet{Bickley22}. This represents our purest sample of mergers, but is still a lower limit of the total post-mergers in the populations for two reasons. First, only galaxies assigned a post-merger probability of $p > 0.75$ were visually inspected and there may be some true post-merger galaxies that did not meet that threshold from the CNN. Second, the quality of imaging, orientation of the galaxy, and timescale of the merger features relative to that of post-starburst features may also cause galaxy mergers to be missed (this is discussed in more detail in Section \ref{IQ}). However, the latter would affect the classification of mergers in both the PSB and control samples proportionally. Thus, while the merger fractions may be lower limits, the excess over the controls is more robust. With this in mind, the visually-confirmed CNN-predicted merger fraction of PCA PSBs is $f_\text{CNN, VC} = (7\pm1)\%$ with an excess over their controls of $Q_\text{CNN, VC} = 29\pm9$. Comparatively, the merger fraction of the E+A PSBs is $f_\text{CNN, VC} = (15\pm3)\%$ with an excess over their controls of $Q_\text{CNN, VC} = 46\pm22$.
    
    \subsection{Visual Classification}
    \label{VCRes}
    
    The identification of mergers by visual classification is not infallible and certainly subjective, but still provides a good comparison to previous studies and a sanity check for our other metrics. Hence, both the E+A and PCA PSB samples in their entirety were visually classified following the strategy outlined in Section \ref{VCmeth}. Because the control samples are ten times larger than the PSB samples only the best-matched control of each PSB are classified for comparison. 147 of the 533 PCA PSBs are visually classified as having disturbed morphologies indicative of a recent merger or interaction giving a merger fraction of $f_\text{VC} = (28\pm2$)\% . In contrast, only 16 of the 533 best-match star-forming controls are identified as mergers, translating to an excess in the PCA PSBs of $Q_\text{VC} = 9.6\pm0.3$. On the other hand, 66/157 ($42\%\pm4$\%) E+A PSBs are identified as post-merger galaxies, in excess of their controls by a factor of $Q_\text{VC} = 9.9\pm0.4$. Thus, from directly inspecting the galaxy images of the PSB and control samples we find a significantly higher fraction of recent mergers in the E+A PSBs than in the PCA PSBs but when compared to their respective control samples, their excesses are statistically equivalent. This result is consistent with the findings from the non-parametric morphology methods (see Section \ref{ShapeAsymmetry}) and from the CNN post-merger classifications (see Section \ref{CNNRes}) and therefore insensitive to the merger identification method.
    
\section{The Effect of Image Quality}
    \label{IQ}

    The depth, atmospheric blurring, and the on-sky pixel scale resolution of the imaging can dramatically affect the non-parametric morphology values assigned to a particular galaxy. The effect is so pernicious that the same galaxy imaged with two different observing programmes in the same bandpass can have vastly different morphology measurements. In this section, we explore the effect of image quality on our non-parametric morphologies and their associated merger fractions with a particular emphasis on comparing the image quality of CFIS to that of SDSS.
    
    \subsection{Testing Variable Image Quality with IllustrisTNG}
    \label{IQIll}
    
    To test the sensitivity of S(G, M$_{20}$), $A$, and $A_S$\footnote{We do not test the sensitivity of the CNN post-merger prediction to variable image quality since it is trained specifically for CFIS-like images. Reliably testing a CNN's ability to recover post-mergers at different image qualities would require re-training the CNN at every image quality tested.} to the quality of imaging (including quality superior to that of CFIS) from which they are derived, we apply our morphology analysis to synthetic galaxy images from the IllustrisTNG simulation . Galaxies from the simulation have no sky noise or atmospheric blurring and have a higher resolution than most imaging systems. From a pristine state free from observational effects, the images can be degraded to varying levels of image quality, including that of CFIS and SDSS, and the variability of the derived morphologies can be tested accordingly.
    
    Synthetic galaxy images are generated from the stellar mass distributions of galaxies in the simulation, used here as a proxy for its stellar light profile. Each synthetic image is generated at the same field of view as the real CFIS images used in this work and at an "observed" redshift of 0.05. To simulate atmospheric blurring and sky noise, the galaxy profile is convolved with a 2-D Gaussian function with varying FWHMs and is co-added with values drawn from a normal distribution with varying standard deviations. The images are then binned to the pixel scale of CFIS. For a single galaxy, a range of FWHMs from 0.1" to 1.7" and depths from 23.5 mag/arcsec$^2$ to 25.5 mag/arcsec$^2$ are combined to form a 9$\times$9 grid of synthetic images with varying quality. An additional 82$^\text{nd}$ image is generated at a depth of 30 mag/arcsec$^2$ and without any blurring effects as a "ground truth" comparison. Each of the 82 images are run through \texttt{statmorph} as outlined in Section \ref{SMimp}.
    
     \begin{figure}
        \centering
        \includegraphics[width=1\linewidth]{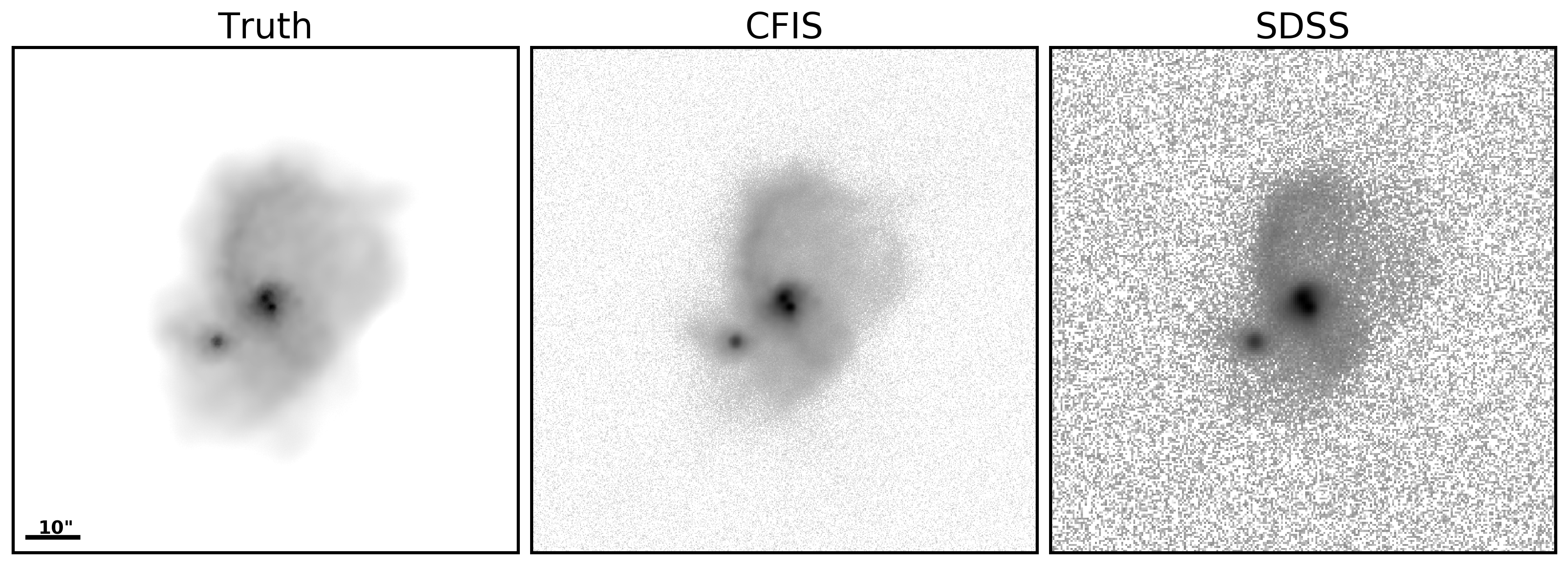}
        \caption{Example synthetic images for the archetypal post-merger galaxy from the IllustrisTNG simulation described in Section \ref{IQIll}. The left panel is the "ground truth" image with no PSF blurring and minimal sky noise. The mock CFIS and SDSS images (middle and right panels) are coadded with sky noise, convolved with a PSF and binned to the pixel scale appropriate for each survey. These synthetic SDSS and CFIS images represent the orange and blue crosses in Figure \ref{SimExample} and are used in the experiment presented in Section \ref{mfracILL}. All three images are normalized and displayed on equivalent logarithmic scales.}
        \label{MockIms}
    \end{figure}
    
\begin{figure}
        \centering
        \includegraphics[width=0.97\linewidth]{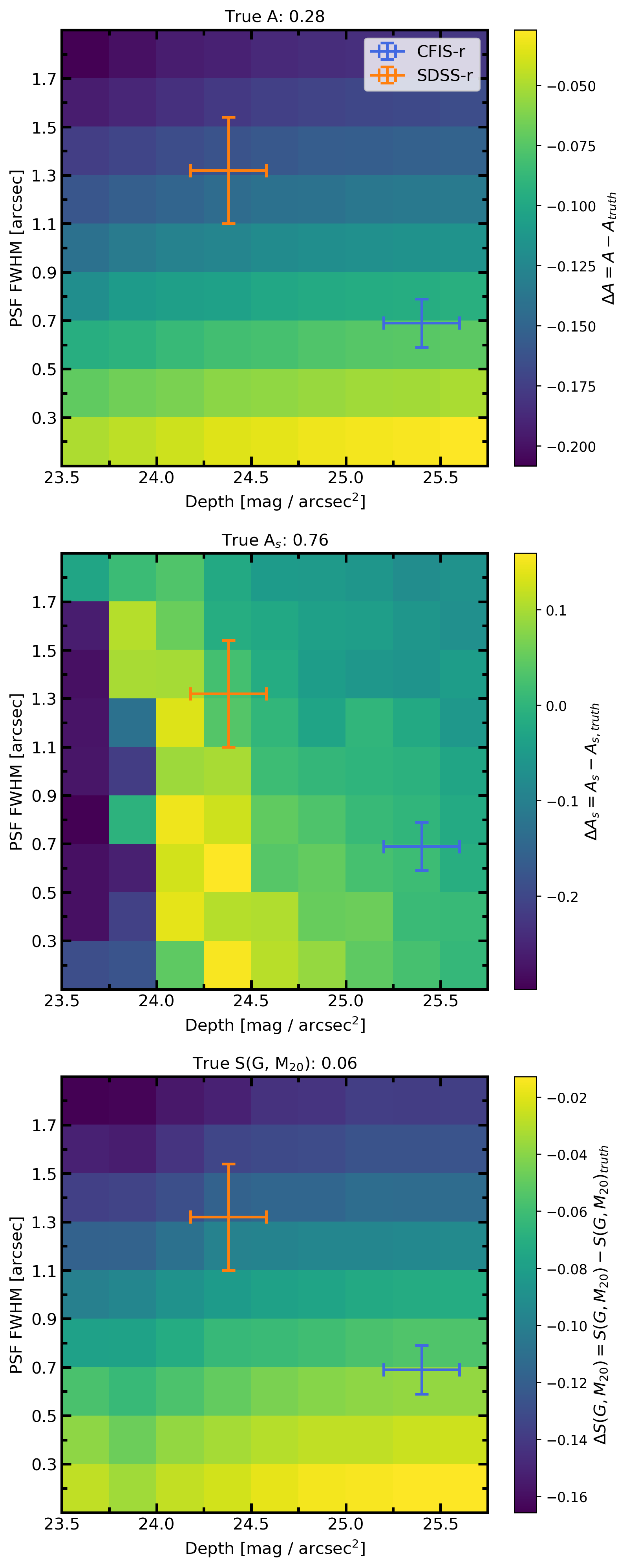}
        \caption{The asymmetry (top), shape asymmetry (middle) and Gini-M$_{20}$ merger statistic (bottom) of a simulated post-merger from IllustrisTNG convolved with a Gaussian PSF with varying FWHM and coadded with sky noise at various depths. Each grid is coloured by the resulting difference from the ground truth morphology of the galaxy, as defined at a depth of 30 mag/arcsec$^2$ and no PSF blurring. Orange and blue crosses portray typical image quality of SDSS and CFIS, respectively. All images are binned to the pixel scale of CFIS: 0.187 arcsec/pixel.}
        \label{SimExample}
    \end{figure}

    Since we expect the dependence of non-parametric morphology metrics on depth and atmospheric blurring to differ between galaxies, we present a case study of one archetypal post-merger galaxy in IllustrisTNG. The $3.51\times10^{10} \text{ M}_\odot$ galaxy in question is at $z = 0.33$ of the simulation and has undergone a merger with a mass ratio of $\mu = 0.553$ within the last snapshot of the simulation (t$_\text{post-merger} \lesssim 162$ Myr). In Figure \ref{MockIms}, we display the "ground truth" image of this archetypal galaxy with minimal realism applied as well as the degraded synthetic images at the image quality CFIS and SDSS. Several late-stage merger features are present including faint, extended, asymmetric tidal tails, a double-nucleus, and a diffuse stellar halo. As a result, the ground truth shape asymmetry (A$_S$ $\thicksim$ 0.76) and Gini-M$_{20}$ merger statistic (S(G, M$_{20}$) $\thicksim$ 0.06) are each well above their respective merger thresholds and thus the galaxy would be classified as a merger by both metrics. However, the ground truth asymmetry is only moderately high (A $\thicksim$ 0.28) placing it below the \citet{Conselice03} merger threshold of 0.35. In Figure \ref{SimExample}, we show how each of the morphology metrics deviate from the ground truth as the image quality varies across the 9$\times$9 grid of synthetic images. 
    
    The top panel of Figure \ref{SimExample} shows that the calculated asymmetry value decreases from the ground truth by a maximum of $-0.208$ at the lowest quality imaging. However, at the maximal depth and minimal atmospheric blurring considered, the difference from the ground truth is minimized to less than 0.003. The diagonal trend indicates that asymmetry tends to be affected equally by atmospheric blurring and image depth, consistent with asymmetry being sensitive to both asymmetric tidal features and high spacial frequency internal disturbances \citep{Saz21}. For additional context, Figure \ref{SimExample} indicates the typical image quality of CFIS and SDSS images with the blue and orange crosses, respectively. For this post-merger galaxy, the asymmetry in CFIS is $\thicksim$0.1 higher (36\% closer to the true value) than if it were measured using SDSS imaging. However, this post-merger galaxy would not be considered a merger in either of the surveys as the ground truth galaxy profile is simply not asymmetric enough to surpass the merger threshold. 
    
    In the middle panel of Figure \ref{SimExample}, we repeat the comparison of true and observed metrics, but now for shape asymmetry. The left-to-right trend evident in the deviation from ground truth of the shape asymmetry indicates that the variation in the shape asymmetry is driven largely by the depth of the imaging rather than the atmospheric blurring. This is consistent with shape asymmetry being sensitive to asymmetric low surface brightness features \citep{Pawlik16} and neglecting internal disturbances. As the depth of the imaging improves, the shape asymmetry increases to a value $\thicksim$0.16 \emph{higher} than the ground truth before decreasing again to be within 0.02 of the ground truth, on average, in the column of highest depth images. As a result, the shape asymmetry of this galaxy would be higher in SDSS than in CFIS. In this case, the shape asymmetry decreases at higher depths because the binary segmentation map begins to include the ultra faint diffuse halo which happens to be more symmetric than the less faint tidal tails. Nonetheless, this galaxy's shape asymmetry exceeds the nominal merger threshold of 0.4 and thus would be classified as a merger regardless of the image quality used in our experiment.
    
    Finally, in the lower panel of Figure \ref{SimExample}, we assess the impact of image quality on S(G, M$_{20})$. In contrast to shape asymmetry, S(G, M$_{20})$ changes dramatically with atmospheric blurring but is less affected by a change in image depth. We attribute this to atmospheric blurring distributing the flux of the bright nuclei over more pixels, reducing G and increasing M$_{20}$, and the additional depth having little influence on the brightest pixels in the image. In the range of CFIS image quality, this galaxy would be marginally considered a merger, but in SDSS the image has been degraded to a degree that it would not be classified as a merger.
    
    With this test, we have shown that the sensitivity of S(G, M$_{20}$), $A$, and $A_S$ to the depth and atmospheric blurring of the imaging differs between each metric. However, we have repeated this experiment on many galaxies and found that the magnitude of each metric's sensitivity depends on the morphological features inherent to the galaxy in question. For example, the shape asymmetry of a galaxy with asymmetric tidal features is expected to increase with increased depth as more asymmetric features are detected, while a galaxy without such features will exhibit minimal change in its shape asymmetry with increased depth \citep[see also][]{Pawlik16}. 
    
    \subsection{The Fraction of Undetected Mergers in CFIS and SDSS}
    \label{mfracILL}
    
    In the previous subsection, we have established that a post-merger galaxy may be given a different merger classification depending on the morphology statistic implemented and the quality of the imaging used. In the case described in Section \ref{IQIll}, the asymmetry suggests the example post-merger galaxy is a non-merger at both CFIS and SDSS image quality, the shape asymmetry indicates it is a recent post-merger for the entire range of image qualities, and the Gini-M$_{20}$ merger statistic classified it as a merger in CFIS imaging but not SDSS. We now assess how image quality representative of SDSS and CFIS will affect our ability to identify mergers from a statistical sample of simulated post-merger galaxies.
    
    We use a sample of 2332 recent IllustrisTNG post-mergers at $z<1$ of the simulation. To be considered, the merger must have been significant ($\mu > 0.1$) and the post-merger remnant galaxy must have a mass between 10$^{10}\text{ M}_\odot$ and 10$^{12}\text{ M}_\odot$. Each post-merger is observed from four distinct viewing angles; each viewing angle provides a new stellar light distribution of a post-merger event, effectively enhancing the post-merger sample by a factor of four to 9328. The 9328 post-merger images are scaled in angular size as if they were at an observed redshift of 0.05 and then degraded to the pixel scale resolution and typical atmospheric blurring and depth of CFIS and SDSS (see Section \ref{data}) before being processed by \texttt{statmorph} as described in Section \ref{SMimp}. 7971 of these galaxies have flag-free morphologies for both the mock SDSS and CFIS images. 
    
    With 7971 mock-observed post-mergers in hand, we can assess the ability of $A$, $A_S$ and S(G, M$_{20}$) to identify recent post-merger galaxies using merger thresholds of 0.35, 0.4 and 0.0, respectively. For each of the three metrics, the number of simulated post-mergers that would be identified as such is shown in Table \ref{TAB2}. Only 6.8\% of the post-merger sample was identified as a merger based on their mock-observed CFIS asymmetries. However, this is an improvement over the 2.5\% that would have been identified using SDSS imaging. S(G, M$_{20}$) recovers 30\% of post-mergers when using CFIS-realistic imaging, an improvement over the 21\% recovered using SDSS imaging. Shape asymmetry appears to be equally effective as S(G, M$_{20}$) at identifying recent mergers, identifying 30\% with CFIS-realistic imaging and 28\% with SDSS-realistic imaging. 
    
    CFIS imaging allows for a greater number of post-mergers to be detected using non-parametric morphologies. However, \emph{more than 70\% of galaxies that have undergone a merger within the last $\thicksim$200 Myr do not exhibit the features necessary to be detected using non-parametric morphology merger indicators}. By the time a post-merger galaxy that goes on to rapidly reduce its star formation rate and would be considered a post-starburst ($\gtrsim$ 500 Myr later), the merger features will likely have faded further and even fewer post-mergers would be recovered. Thus, based on the results presented in Section \ref{QuantMorph}, we cannot exclude the possibility that the entire PSB sample is of merger origin.
    
    \begin{table}
    \centering
    \begin{tabular}{|l|c|c|c|}
    \hline
     & A $>0.35$ & A$_S > 0.4$ & S(G, M$_{20}$) $> 0$ \\
    \hline
    
    SDSS realism & 2.5\% & 28\% & 21\%\\
    
    CFIS realism & 6.8\% & 30\% & 30\%\\
    
    \hline
    \end{tabular}
    \caption{The fraction of recent post-merger galaxies from the IllustrisTNG simulation that would be detected as a merger using automated non-parametric morphology metrics. Realistic observational effects such as atmospheric blurring, pixel scale resolution and sky noise tuned to typical CFIS and SDSS imaging are applied to the synthetic images of 7971 simulated post-mergers.}
    \label{TAB2}
    \end{table}

\subsection{The Difference in PSB Merger Fraction in CFIS and SDSS}
\label{mfrac_cvs}

    It is clear from our tests with IllustrisTNG that the quality of imaging has a significant effect on the computed non-parametric morphologies and thus the number of mergers detected. We now return to our observed datasets and repeat our morphological analysis using SDSS imaging to understand how the improved image quality offered by CFIS changes the non-parametric morphologies of galaxies and how that difference alters the PSB merger fraction results from Section \ref{Results}.
    
    The 168,477 galaxies in CFIS DR2 and SDSS DR7 were processed with \texttt{statmorph} again, but with SDSS \emph{r}-band images instead of CFIS. The implementation is identical to the description in Section \ref{SMimp} except for minor changes to the treatment of the weightmap and segmentation map. Weightmaps are not available for SDSS images as they are for CFIS, so a gain is provided to \texttt{statmorph} instead. Using a gain or weightmap does not have any effect on the non-parametric morphologies computed by \texttt{statmorph}. The segmentation map was generated the same way but in some cases the target galaxy was too faint in SDSS to be detected by Source Extractor. In such cases, the detection threshold was reduced from 1.5$\sigma$ to 1.1$\sigma$, and if the galaxy was still too faint to be detected, it was discarded from the catalogue. The difference in derived morphologies when using CFIS and SDSS imaging for all $\thicksim$170,000 galaxies in the overlap between the surveys is described in detail in Appendix \ref{appA}.
    
    We re-analyze the merger fraction of PSBs with the three non-parametric morphology metrics derived from SDSS and by visual classification of the SDSS images. Re-training and re-running a CNN to identify mergers in SDSS would be a significant undertaking that is outside the scope of this work. Furthermore, we restrict our analysis here to only the PCA PSBs since it is our intention to compare the effect of image quality on our results, not PSB selection strategy. We use the PCA PSB sample instead of the E+A PSB sample simply because it is larger and allows for more statistically robust conclusions to be drawn. When selecting the PSB and control samples in Section \ref{ss}, we required galaxies to have flag-free CFIS morphologies. For a fair comparison between CFIS and SDSS galaxies, we require that every PSB and its controls have both flag-free CFIS and SDSS morphologies. This limits the PCA PSB sample size to 488. 
    
    \begin{figure}
        \centering
        \includegraphics[width=1\linewidth]{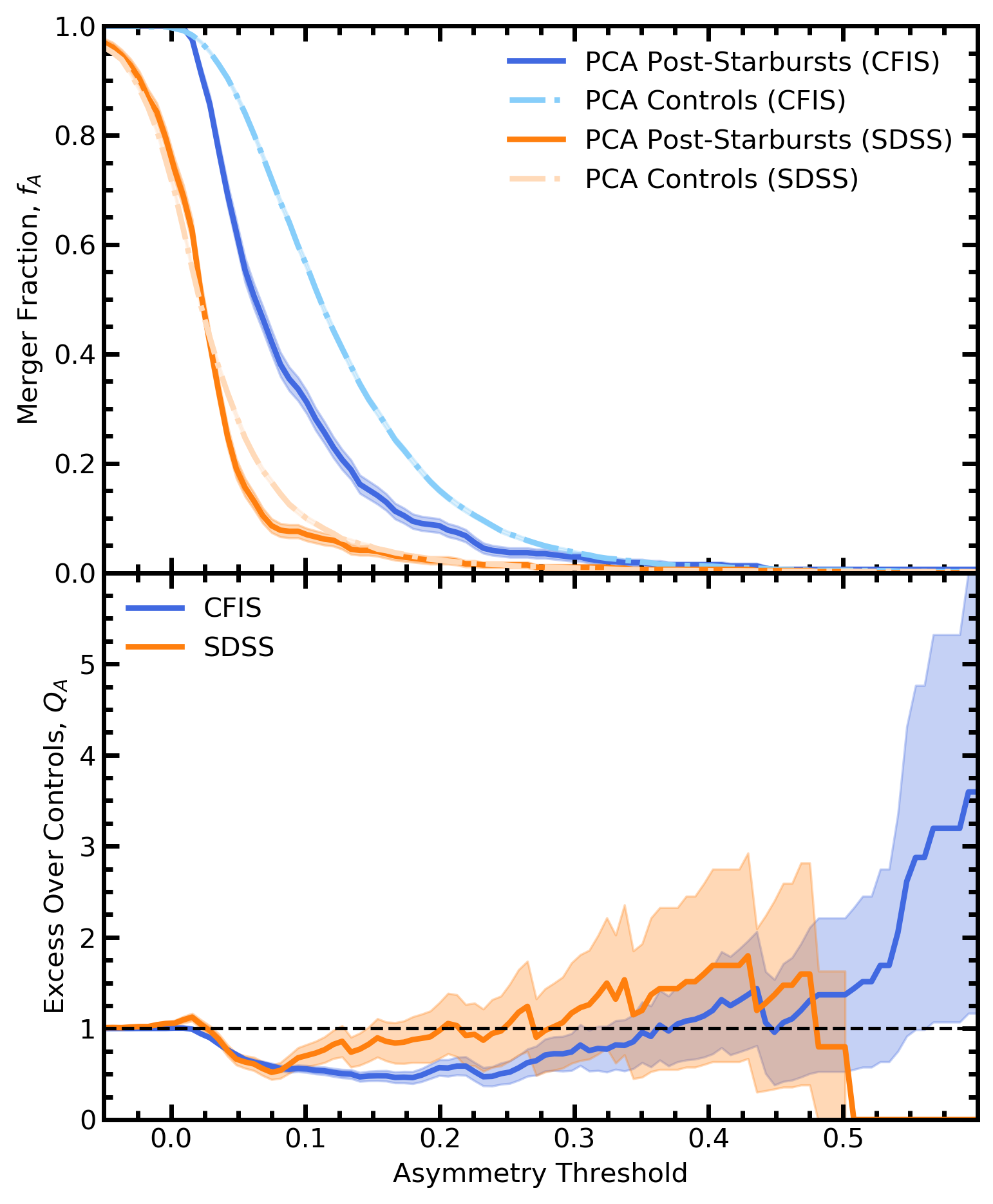}
        \caption{The same as Figure \ref{DRA} but considering a subset of the PCA PSBs and their star-forming controls defined by each PSB and its matched control having flag-free morphologies for both CFIS and SDSS imaging. The orange curves show how our results would differ if we were to use SDSS imaging instead of CFIS.}
        \label{SDSS_DRA}
    \end{figure}

    In Sections \ref{IQIll} and \ref{mfracILL} we found that for recent mergers from the IllustrisTNG simulation, the image quality can have a significant effect on the derived value of S(G, M$_{20}$) for post-merger galaxies and on the number of identified mergers using the same metric; a similar effect is observed for the PSB sample. Recall that in Section \ref{GM20-Results}, we found that the S(G, M$_{20}$))-derived merger fraction for the full sample of 533 PSBs was (8$\pm$1)\%, a factor of 1.8$\pm$0.3 over their controls. Once the PSBs and their controls with SDSS morphology flags are removed, the merger fraction of the 488 remaining PSBs with CFIS imaging changes to (7$\pm$1)\%, a factor of 2.3$\pm$0.7 more than their controls. Using SDSS imaging to determine the merger fraction of the same 488 PSBs reduces the merger fraction to (6$\pm$1)\%, but still a factor of 1.7$\pm$0.4 more than their controls. Thus, fewer mergers are identified in the same sample but within statistical error the same S(G, M$_{20}$) results derived using CFIS imaging would be recovered using SDSS imaging.

    In general, the asymmetry of galaxies in CFIS tend to be higher than in SDSS (see Appendix \ref{appA}). For the sample of 488 PSBs, CFIS imaging gives a merger fraction of (1.8$\pm$0.6)\%, translating to an excess of (0.9$\pm$0.3) over their controls at a threshold of $A>0.35$. In SDSS, the merger fraction of PSBs is (0.6$\pm$0.4)\%, three times less than in CFIS but translating to a statistically equivalent excess over their controls of (1.2$\pm$0.7).

    \begin{figure}
        \centering
        \includegraphics[width=1\linewidth]{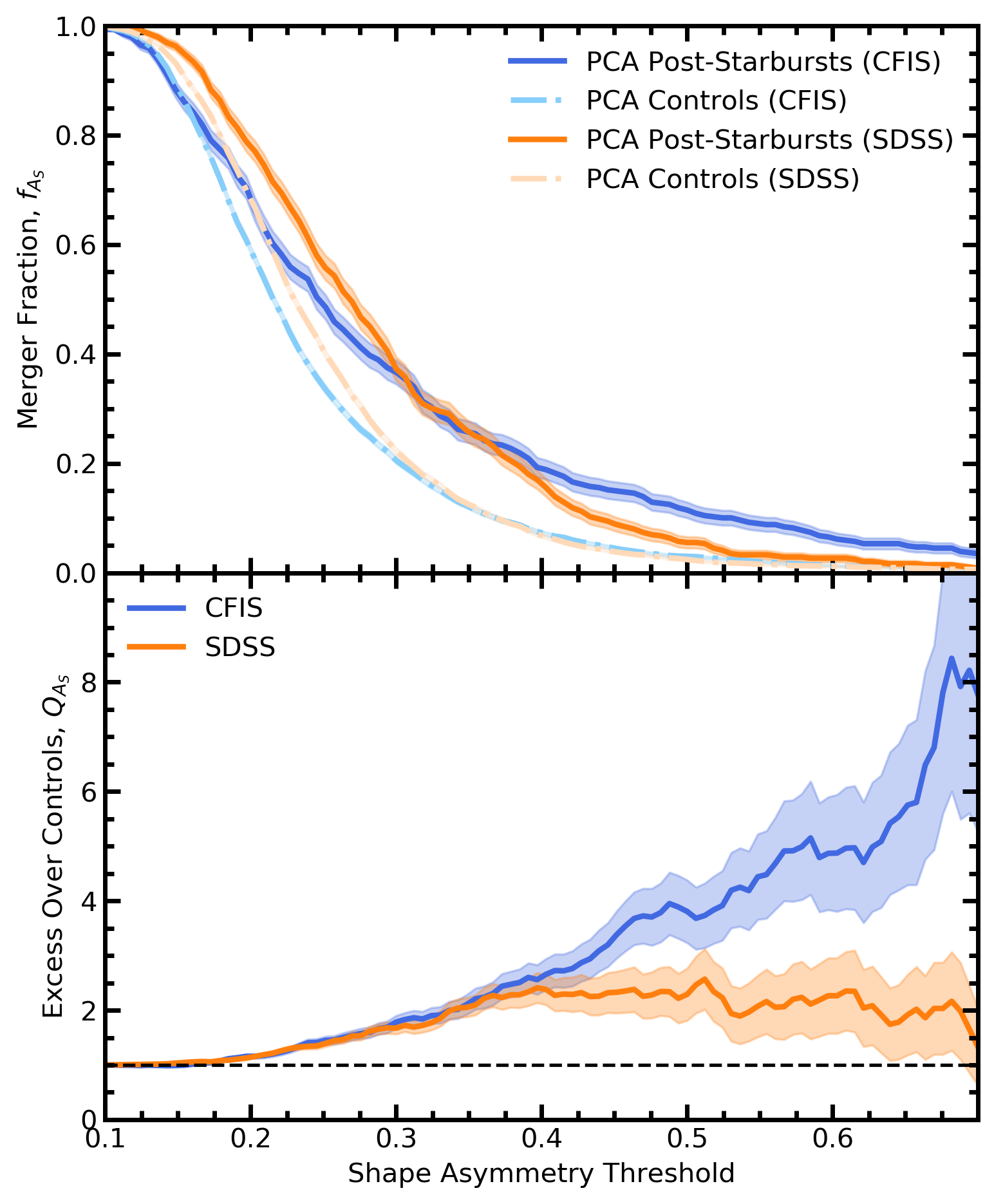}
        \caption{The same as Figure \ref{AsDR} but considering a subset of the PCA PSBs and their star-forming controls defined by each PSB and its matched control having flag-free morphologies for both CFIS and SDSS imaging. The orange curves show how our results would differ if we were to use SDSS imaging instead of CFIS.}
        \label{SDSS_DRAs}
    \end{figure}
    
    In Figure \ref{SDSS_DRA}, we test how the merger fraction and excess would change in CFIS and SDSS using various asymmetry thresholds to identify mergers. In the top panel of Figure \ref{SDSS_DRA} we see that regardless of the asymmetry threshold used, there are fewer identified mergers in SDSS than in CFIS. While in CFIS PSBs have reduced asymmetries and thus fewer identified mergers than their controls at intermediate asymmetry thresholds, the same is not true for SDSS because the poorer image quality already dictates lower asymmetries in both samples. This drives the excess in SDSS to unity for almost all considered thresholds. In CFIS, there are a small number of galaxies with very high asymmetries ($A>0.5$) which generate an excess at high thresholds. These are not present in the SDSS sample and since there are no galaxies in either the PSB or control samples with high asymmetries, the excess becomes incalculable.
    
    \begin{figure*}
        \centering
        \includegraphics[width=\linewidth]{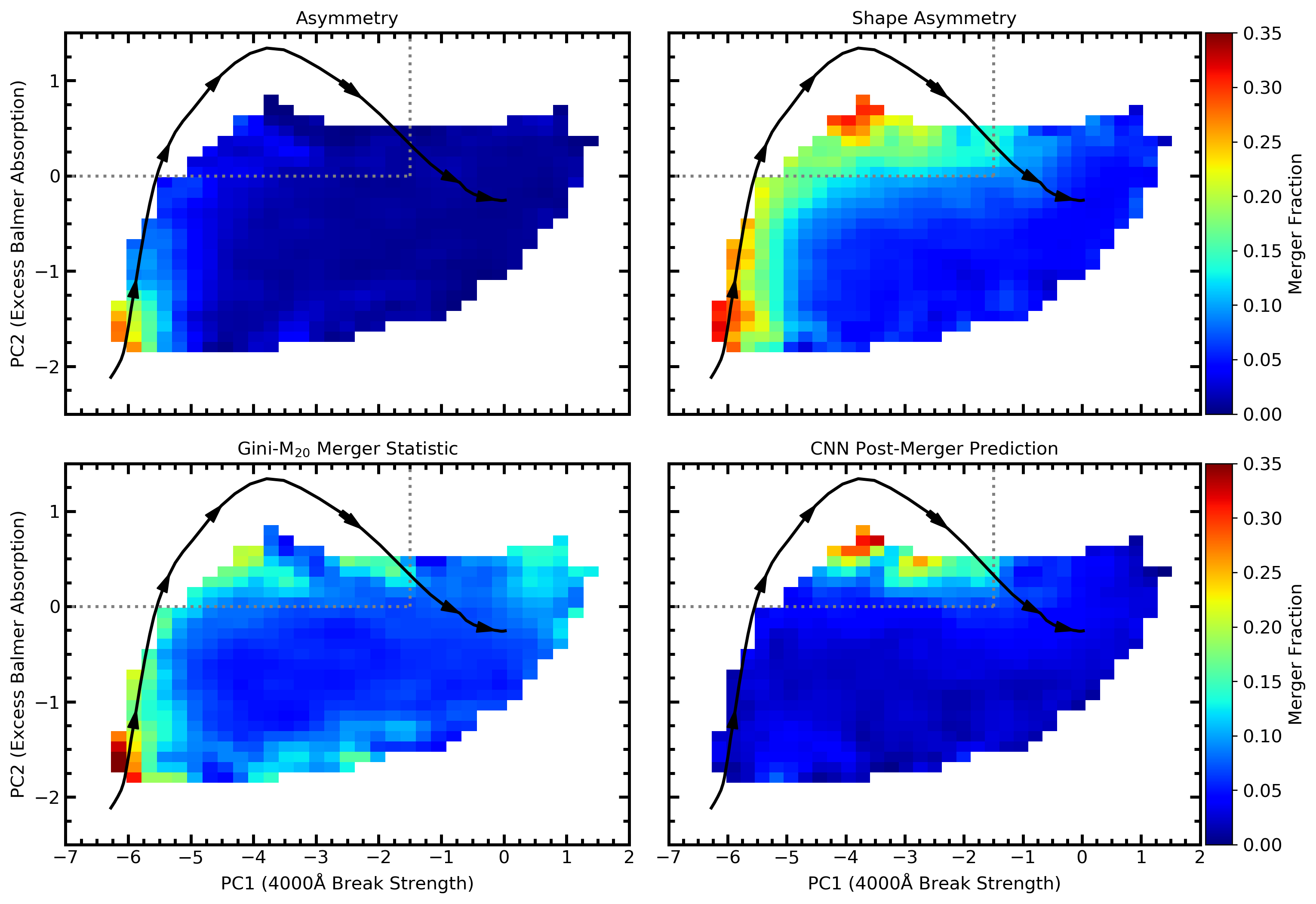}
        \caption{The merger fraction, as determined using the default threshold of four merger identification metrics, throughout the entire PC1-PC2 plane. The merger fraction is determined in each bin and then smoothed by a 3$\times$3 kernel. The colour-merger fraction relationship is equal for each of the four panels, as quantified by the colour bar on the right. Over-plotted in black is a model evolutionary track of a tophat starburst of 0.03 Gyr duration and burst mass fraction of 3\% from \citet{Wild07}. The track begins 35 Myr after the onset of the starburst (5 Myr after it ends) and the arrows are placed 0.05, 0.1, 0.2, 0.5, 1 and 2 Gyr after the onset of the burst. The dotted grey line indicates the PCA PSB selection region described in Section \ref{ss}.}
        \label{PCPlaneMF}
    \end{figure*}
    
    Adopting a merger threshold of $A_S > 0.4$, the CFIS merger fraction is (19$\pm$2)\%, a factor of 2.6$\pm$0.3 over the controls. In SDSS, the merger fraction is only marginally lower at (16$\pm$2)\%, in excess of their controls by a factor of 2.3$\pm$0.3. Thus, at the default merger threshold, the same conclusion would be drawn using either survey. It is only at higher and lower merger thresholds that the difference in quality of imaging offered by each survey makes a difference in the conclusion, as shown in Figure \ref{SDSS_DRAs}. At lower thresholds, the  merger fraction is higher in SDSS because galaxies with inherently low asymmetry have higher shape asymmetry in SDSS than CFIS due to noise and resolution effects (see Appendix \ref{appA}). The noise and resolution affect PSBs and controls equally and so there is no difference between the excess merger fraction at low thresholds. At higher merger thresholds, the excess over the controls in CFIS continues to increase to $Q\thicksim8$. In SDSS, the shallower imaging causes fewer PSBs to have very high shape asymmetries and thus the excess over the controls stagnates at $Q\thicksim2$. Therefore, whilst the merger excesses are similar for SDSS and CFIS images when asymmetry is used (Figure \ref{SDSS_DRA}), shape asymmetry finds a higher merger excess in the deeper imaging.

    Each PSB and best-match star-forming control were classified by eye as outlined in Section \ref{VCmeth}. Since visual classification does not rely on \texttt{statmorph} morphologies, we return to the full sample of 533 PCA PSBs. Recall from Section \ref{VCRes} that the visual classification of the PCA PSBs with CFIS images gave a merger fraction of ($28\pm2$)\% with an excess over their controls of $Q_\text{VC} = 9.6\pm0.3$. Since SDSS has shallower and lower resolution imaging, there are many cases where merger features are too faint to be seen. As a result, classification with SDSS imaging truncates the number of identified mergers resulting in a merger fraction of ($17\pm2$)\% and an excess of $5.7\pm0.3$
    
    The tests presented in Section \ref{IQ} have shown that all metrics identify incomplete numbers of mergers and are sensitive to the depth and resolution such that more mergers are identified in deeper and higher resolution imaging. These factors are likely to be one of the main reasons for the diversity of results quoted in the literature (see Table \ref{MergerSummary}). The use of CFIS imaging allows us to identify a more reliable merger fraction than previous studies that use SDSS imaging. However, our results also indicate that even higher quality imaging than CFIS would extend the number of identifiable mergers. The merger fractions derived from CFIS imaging are therefore likely to be a lower limit of the true merger fraction of PSBs.

\section{Discussion of PSB and Merger Timescales}
\label{disc}
\label{Efficacy}

    In Section \ref{Results}, we quantified the merger fraction of PSBs using a variety of merger identification methods, each with varying minimum thresholds above which a galaxy is classified as a merger. We have often referred to the fact that if the initial starburst was caused by a merger, then its post-merger features will fade (perhaps beyond detection) by the time the galaxy becomes a post-starburst. However, our analysis uses strict post-starburst definitions and thus does not provide any insight into how the number of identified mergers would change in an evolutionary epoch after the initial starburst but before the galaxy becomes a post-starburst by our definitions. In Figure \ref{PCPlaneMF}, we present a retrospective assessment of the merger fractions computed using the default thresholds of several merger identification metrics throughout the entire PC1-PC2 plane. To provide context for the evolution of a (post-)starburst galaxy through the plane, we have overlaid a model evolutionary track of a starburst of 0.03 Gyr duration and burst mass fraction of 3\% (SFR$_\text{burst} = 43 \text{ M}_\odot / \text{yr}$) superimposed upon a quiescent galaxy spectrum from \citet{Wild07}.
    
    With the additional context of the PC1-PC2 plane, the merger fractions from Section \ref{Results} and the efficacy of each merger identification method used to assess the merger fraction of PSBs can be better understood. Asymmetry (top left panel of Figure 12) identifies significant numbers of starburst galaxies (located in the lower left corner of the plane, where the starburst track begins) as mergers. However, the number of identified mergers using asymmetry decreases rapidly as the starburst evolves and the merger features begin to fade. By the time a starburst galaxy reaches the post-starburst phase, as indicated by the region bounded by the grey dotted lines, very few galaxies remain detectable as mergers based on their asymmetry calculation, consistent with the low merger fraction measured in Section \ref{Asymmetry}.
    
    The Gini-M$_{20}$ merger statistic (lower left panel of Figure 12) behaves very similarly but persists further into the post-starburst phase than asymmetry. However, it should also be noted that the merger fraction remains at $\thicksim$5\%-10\% throughout the PC1-PC2 plane indicating a significant number of false-positive merger predictions. Both observations of the Gini-M$_{20}$ merger fractions in the PC1-PC2 plane are consistent with the results found in Section \ref{GM20-Results}. Furthermore, both asymmetry and the Gini-M$_{20}$ merger statistic are confirmed to be most effective at identifying interacting or very recent post-merger galaxies, but less so for the late-stage post-mergers expected in the post-starburst phase.
    
    Shape asymmetry (top right panel of Figure 12) identifies significant fractions of mergers in recent starbursts but, unlike asymmetry, the information persists well into the post-starburst phase. However, shape asymmetry at its default threshold identifies $\thicksim$5\% of galaxies as mergers throughout the PC1-PC2 plane, perhaps indicating a large number of false-positive contamination. 
    
    The CNN (lower right panel of Figure 12) does not predict a high merger fraction in the starburst region of the PC1-PC2 plane because it was trained to specifically find post-merger galaxies and to exclude interacting galaxies and ongoing pre-coalescence mergers. The region where the CNN predicts a fraction of post-merger galaxies $\gtrsim$10\% is approximately the same as the region from which we selected our PSBs. Remarkably, this is the only region in the plane where this is the case indicating that a very significant fraction of CNN-predicted post-mergers in the overlap of CFIS and SDSS are post-starburst galaxies \citep[see][]{Ellison22}.

    The panels in Figure \ref{PCPlaneMF} that use non-parametric morphology statistics as merger indicators suggest that the recovered merger fraction immediately after the starburst event may be as high as $\thicksim$30\%. This merger fraction decreases significantly in the cases of asymmetry and the Gini-M$_{20}$ merger statistic as the model post-starburst track progresses and merger features fade. However, shape asymmetry and the CNN post-merger prediction maintain a merger fraction as high as $\thicksim$30\% in several bins within the post-starburst regime, indicating they are the more reliable merger identification methods within the context of PSBs. Recall that in Section \ref{mfracILL}, we found that only $\thicksim$30\% of very recent post-mergers in IllustrisTNG would be recovered using the same metrics. Thus, it remains an open question whether the majority of PSBs in our sample which are not identified as mergers are caused by mergers with rapidly fading merger features that are undetected by our metrics or because other non-merger mechanisms are generating (post-)starburst galaxies.

\section{Summary}
\label{Summary}

    \begin{figure*}
        \centering
        \includegraphics[width=1\linewidth]{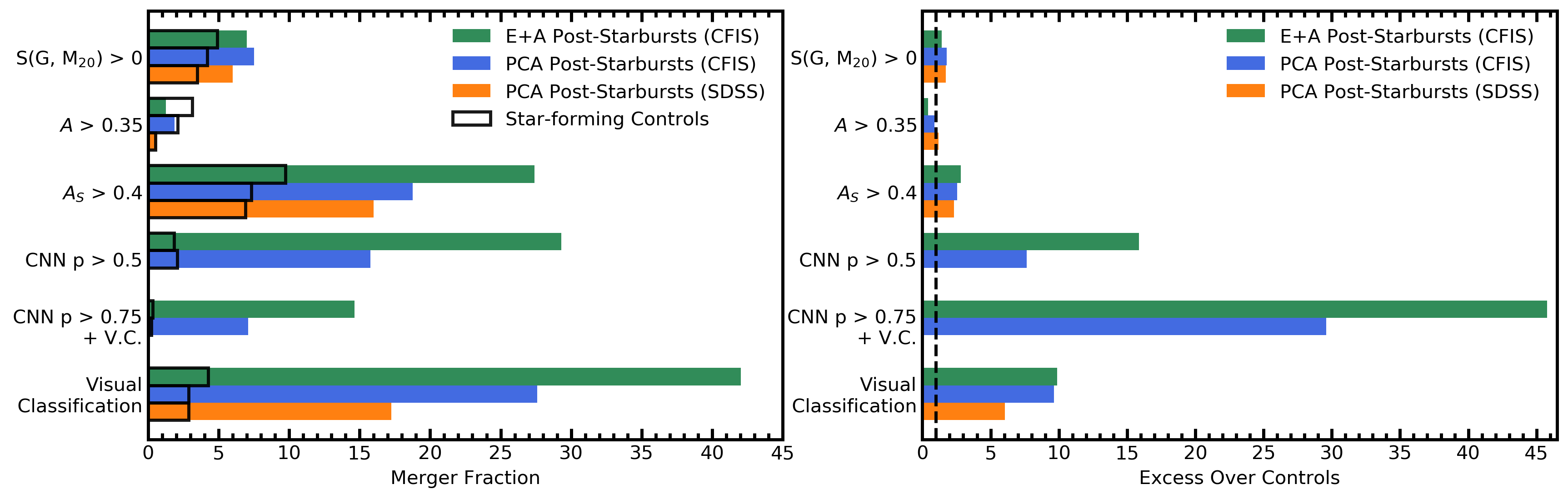}
        \caption{\emph{Left}: A visual summary of the calculated merger fractions for a range of merger definitions, PSB selection methods, and image quality. While each coloured bar represents the PSB merger fraction, the black line gives the merger fraction of the matched star-forming controls. \emph{Right}: A visual summary of the excess merger fraction in the PSB samples over their star-forming controls for a range of merger definitions, PSB selection methods, and image quality. The dashed black line indicates proportional mergers in the PSBs and their controls. \emph{Both}: Each merger identification method is subject to a different interpretation: S(G, M$_{20}$) is best for identifying only very recent mergers and has a significant non-merger contamination; asymmetry is not sensitive to faint asymmetric features and is thus only sensitive to recent and extreme mergers; shape asymmetry is more sensitive to faint asymmetric features and thus detects more mergers in the PSB sample than standard asymmetry; the default CNN threshold identifies a more reliable sample of post-mergers with fewer non-merger contaminants than other automated methods; the visually confirmed CNN sample has near-perfect purity but only identifies a subset of all mergers; visual classification may be the truest representation of the merger fraction of these samples but is a subjective process that will not be applicable to large datasets.}
        \label{SummaryFig}
    \end{figure*}

    Previous studies have found that significant fractions of post-starburst galaxies have disturbed morphologies indicative of a recent merger or interaction with another galaxy \citep[e.g.][]{Zab96,Blake04,Yang08,Pawlik18}. However, poor sample statistics, low quality imaging, and inconsistent methods of both selecting PSBs and identifying mergers contribute to a wide spread of reported merger fractions (see Table \ref{MergerSummary}) and restrict the field from converging on a quantitative understanding of the importance of galaxy mergers to rapid quenching. In this work, we have addressed these issues  by assembling two large samples of PSBs using an "E+A" selection technique \citep{Goto05} and a PCA selection technique \citep{Wild07} from the overlap of SDSS DR7 (which offers spectra for identifying PSBs) and CFIS DR2 (which offers deeper and higher resolution imaging than SDSS). We have assessed the merger fraction of the two samples of PSBs using three merger identification strategies: non-parametric morphology statistics \citep{RG19}, a CNN trained to identify post-merger galaxies in CFIS \citep{Bickley21}, and by visual inspection of the galaxy images. In Figure \ref{SummaryFig}, we present a visual summary of our main conclusions:

    \begin{itemize}
    
        \item \textbf{The merger fraction of a given sample of post-starburst galaxies is subject to change when different merger identification methods are applied.} We find that the merger fraction of PCA PSBs vary from 19\% using shape asymmetry, to 16\% using the CNN, to 28\% when visually classified (see Section \ref{Results} and Figures \ref{AsDR}-\ref{FCNN}).
        
        \item \textbf{The merger fraction of a given sample of post-starburst galaxies is subject to change when a different post-starburst selection strategy is implemented, but the excess number of mergers compared to the control sample remains the same.} This conclusion is ubiquitous for the E+A and PCA PSBs across all three merger identification strategies implemented (see Section \ref{Results} and Figures \ref{DRA}-\ref{FCNN}).
        
        \item \textbf{The merger fraction of a given sample of post-starburst galaxies is subject to change when different quality of imaging is used.} For example, the visually classified merger fraction of PCA PSBs is 28\% when using deep and higher resolution CFIS imaging and only 17\% when using lower quality SDSS imaging (see Section \ref{IQ} and Figures \ref{SDSS_DRA}-\ref{SDSS_DRAs}). 
        
        \item There is a clear excess of mergers in the two samples of post-starbursts studied in this work indicating \textbf{post-starbursts are often caused by mergers} (see Section \ref{Results} and Figures \ref{DRA}-\ref{FCNN}). Since the merger fraction is not 100\%, other mechanisms may be causing galaxies to become post-starbursts. However, our tests with simulated post-mergers have shown that non-parametric morphology metrics only recover $\thicksim$30\% of recent post-mergers (see Section \ref{IQIll} and Table \ref{TAB2}). Thus, we cannot rule out that nearly all PSBs are post-mergers with a majority of PSB merger features having faded beyond detection.
        
    \end{itemize}
    
\section*{Acknowledgements}

SW would like to respectfully acknowledge that the sites on which this paper was written are the ancestral territories of Indigenous peoples. As we explore the shared sky, we acknowledge our responsibilities to honour those who were here before us, and their continuing relationships to these lands. We strive for respectful relationships and partnerships with all the peoples of these lands as we move forward together towards reconciliation and decolonization. SW would also like to thank Fran \& Wayne Roberts and Alan \& Trudie Wilkinson whose support was monumental to this work. CB gratefully acknowledges support from the Natural Sciences and Engineering Council of Canada (NSERC) as part of their post-doctoral fellowship program (PDF-546234-2020) and VW acknowledges STFC grant ST/V000861/1.

This research was enabled by computational resources provided by the Digital Research Alliance of Canada on the Cedar cluster (https://alliancecan.ca/en). Furthermore, this work is based on data obtained as part of the Canada-France Imaging Survey, a CFHT large program of the National Research Council of Canada and the French Centre National de la Recherche Scientifique. Based on observations obtained with MegaPrime/MegaCam, a joint project of CFHT and CEA Saclay, at the Canada-France-Hawaii Telescope (CFHT) which is operated by the National Research Council (NRC) of Canada, the Institut National des Science de l’Univers (INSU) of the Centre National de la Recherche Scientifique (CNRS) of France, and the University of Hawaii. This research used the facilities of the Canadian Astronomy Data Centre operated by the National Research Council of Canada with the support of the Canadian Space Agency. This research is based in part on data collected at Subaru Telescope, which is operated by the National Astronomical Observatory of Japan. We are honored and grateful for the opportunity of observing the Universe from Maunakea, which has the cultural, historical and natural significance in Hawaii. Pan-STARRS is a project of the Institute for Astronomy of the University of Hawaii, and is supported by the NASA SSO Near Earth Observation Program under grants 80NSSC18K0971, NNX14AM74G, NNX12AR65G, NNX13AQ47G, NNX08AR22G, YORPD20\_2-0014 and by the State of Hawaii.

Funding for the SDSS and SDSS-II has been provided by the Alfred P. Sloan Foundation, the Participating Institutions, the National Science Foundation, the U.S. Department of Energy, the National Aeronautics and Space Administration, the Japanese Monbukagakusho, the Max Planck Society, and the Higher Education Funding Council for England. The SDSS Web Site is http://www.sdss.org/. The SDSS is managed by the Astrophysical Research Consortium for the Participating Institutions. The Participating Institutions are the American Museum of Natural History, Astrophysical Institute Potsdam, University of Basel, University of Cambridge, Case Western Reserve University, University of Chicago, Drexel University, Fermilab, the Institute for Advanced Study, the Japan Participation Group, Johns Hopkins University, the Joint Institute for Nuclear Astrophysics, the Kavli Institute for Particle Astrophysics and Cosmology, the Korean Scientist Group, the Chinese Academy of Sciences (LAMOST), Los Alamos National Laboratory, the Max-Planck-Institute for Astronomy (MPIA), the MaxPlanck-Institute for Astrophysics (MPA), New Mexico State University, Ohio State University, University of Pittsburgh, University of Portsmouth, Princeton University, the United States Naval Observatory, and the University of Washington.

\section*{Data Availability}

A subset of the raw data underlying this article are publicly available via the Canadian Astronomical Data Center at http://www.cadc-ccda.hia-iha.nrc-cnrc.gc.ca/en/megapipe/. The remaining raw data and all processed data are available to members of the Canadian and French communities via reasonable requests to the principal investigators of the Canada-France Imaging Survey, Alan McConnachie and Jean-Charles Cuillandre. All data will be publicly available to the international community at the end of the proprietary period, scheduled for 2023.

The morphology catalogue of all galaxies in the overlap of CFIS DR2 and SDSS DR7 is available upon request.

\bibliographystyle{mnras}
\bibliography{Sbib}

\begin{thebibliography}{}
\makeatletter
\relax
\def\mn@urlcharsother{\let\do\@makeother \do\$\do\&\do\#\do\^\do\_\do\%\do\~}
\def\mn@doi{\begingroup\mn@urlcharsother \@ifnextchar [ {\mn@doi@}
  {\mn@doi@[]}}
\def\mn@doi@[#1]#2{\def\@tempa{#1}\ifx\@tempa\@empty \href
  {http://dx.doi.org/#2} {doi:#2}\else \href {http://dx.doi.org/#2} {#1}\fi
  \endgroup}
\def\mn@eprint#1#2{\mn@eprint@#1:#2::\@nil}
\def\mn@eprint@arXiv#1{\href {http://arxiv.org/abs/#1} {{\tt arXiv:#1}}}
\def\mn@eprint@dblp#1{\href {http://dblp.uni-trier.de/rec/bibtex/#1.xml}
  {dblp:#1}}
\def\mn@eprint@#1:#2:#3:#4\@nil{\def\@tempa {#1}\def\@tempb {#2}\def\@tempc
  {#3}\ifx \@tempc \@empty \let \@tempc \@tempb \let \@tempb \@tempa \fi \ifx
  \@tempb \@empty \def\@tempb {arXiv}\fi \@ifundefined
  {mn@eprint@\@tempb}{\@tempb:\@tempc}{\expandafter \expandafter \csname
  mn@eprint@\@tempb\endcsname \expandafter{\@tempc}}}

\bibitem[\protect\citeauthoryear{{Abazajian} et~al.,}{{Abazajian}
  et~al.}{2009}]{SDSS-DR7}
{Abazajian} K.~N.,  et~al., 2009, \mn@doi [\apjs]
  {10.1088/0067-0049/182/2/543}, \href
  {https://ui.adsabs.harvard.edu/abs/2009ApJS..182..543A} {182, 543}

\bibitem[\protect\citeauthoryear{{Alatalo} et~al.,}{{Alatalo}
  et~al.}{2016a}]{Alatalo16-SPOGsampledescription}
{Alatalo} K.,  et~al., 2016a, \mn@doi [\apjs] {10.3847/0067-0049/224/2/38},
  \href {https://ui.adsabs.harvard.edu/abs/2016ApJS..224...38A} {224, 38}

\bibitem[\protect\citeauthoryear{{Alatalo} et~al.,}{{Alatalo}
  et~al.}{2016b}]{Alatalo2016}
{Alatalo} K.,  et~al., 2016b, \mn@doi [\apj] {10.3847/0004-637X/827/2/106},
  \href {https://ui.adsabs.harvard.edu/abs/2016ApJ...827..106A} {827, 106}

\bibitem[\protect\citeauthoryear{{Baldry}, {Glazebrook}, {Brinkmann},
  {Ivezi{\'c}}, {Lupton}, {Nichol}  \& {Szalay}}{{Baldry}
  et~al.}{2004}]{Baldry04}
{Baldry} I.~K.,  {Glazebrook} K.,  {Brinkmann} J.,  {Ivezi{\'c}} {\v{Z}}.,
  {Lupton} R.~H.,  {Nichol} R.~C.,   {Szalay} A.~S.,  2004, \mn@doi [\apj]
  {10.1086/380092}, \href
  {https://ui.adsabs.harvard.edu/abs/2004ApJ...600..681B} {600, 681}

\bibitem[\protect\citeauthoryear{Barbary}{Barbary}{2016}]{Barbary2016}
Barbary K.,  2016, \mn@doi [Journal of Open Source Software]
  {10.21105/joss.00058}, 1, 58

\bibitem[\protect\citeauthoryear{Bekki, Couch, Shioya  \& Vazdekis}{Bekki
  et~al.}{2005}]{Bekki05}
Bekki K.,  Couch W.~J.,  Shioya Y.,   Vazdekis A.,  2005, \mn@doi [\mnras]
  {10.1111/j.1365-2966.2005.08932.x}, 359, 949

\bibitem[\protect\citeauthoryear{{Bell} et~al.,}{{Bell} et~al.}{2012}]{Bell12}
{Bell} E.~F.,  et~al., 2012, \mn@doi [\apj] {10.1088/0004-637X/753/2/167},
  \href {https://ui.adsabs.harvard.edu/abs/2012ApJ...753..167B} {753, 167}

\bibitem[\protect\citeauthoryear{{Bertin} \& {Arnouts}}{{Bertin} \&
  {Arnouts}}{1996}]{BA1996}
{Bertin} E.,  {Arnouts} S.,  1996, \mn@doi [\aaps] {10.1051/aas:1996164}, \href
  {https://ui.adsabs.harvard.edu/abs/1996A&AS..117..393B} {117, 393}

\bibitem[\protect\citeauthoryear{Bickley et~al.,}{Bickley
  et~al.}{2021}]{Bickley21}
Bickley R.~W.,  et~al., 2021, \mn@doi [\mnras] {10.1093/mnras/stab806}, 504,
  372

\bibitem[\protect\citeauthoryear{{Bickley}, {Ellison}, {Patton}, {Bottrell},
  {Gwyn}  \& {Hudson}}{{Bickley} et~al.}{2022}]{Bickley22}
{Bickley} R.~W.,  {Ellison} S.~L.,  {Patton} D.~R.,  {Bottrell} C.,  {Gwyn} S.,
    {Hudson} M.~J.,  2022, \mn@doi [\mnras] {10.1093/mnras/stac1500}, \href
  {https://ui.adsabs.harvard.edu/abs/2022MNRAS.tmp.1474B} {}

\bibitem[\protect\citeauthoryear{{Binney} \& {Tremaine}}{{Binney} \&
  {Tremaine}}{1987}]{BT87}
{Binney} J.,  {Tremaine} S.,  1987, {Galactic dynamics}

\bibitem[\protect\citeauthoryear{Blake et~al.,}{Blake et~al.}{2004}]{Blake04}
Blake C.,  et~al., 2004, \mn@doi [\mnras] {10.1111/j.1365-2966.2004.08351.x},
  355, 713

\bibitem[\protect\citeauthoryear{Bottrell, Simard, Mendel  \& Ellison}{Bottrell
  et~al.}{2019a}]{B19-2}
Bottrell C.,  Simard L.,  Mendel J.~T.,   Ellison S.~L.,  2019a, \mn@doi
  [\mnras] {10.1093/mnras/stz855}, 486, 390

\bibitem[\protect\citeauthoryear{{Bottrell} et~al.,}{{Bottrell}
  et~al.}{2019b}]{Bottrell19CNNReal}
{Bottrell} C.,  et~al., 2019b, \mn@doi [\mnras] {10.1093/mnras/stz2934}, \href
  {https://ui.adsabs.harvard.edu/abs/2019MNRAS.490.5390B} {490, 5390}

\bibitem[\protect\citeauthoryear{{Bottrell}, {Hani}, {Teimoorinia}, {Patton}
  \& {Ellison}}{{Bottrell} et~al.}{2022}]{Bottrell22}
{Bottrell} C.,  {Hani} M.~H.,  {Teimoorinia} H.,  {Patton} D.~R.,   {Ellison}
  S.~L.,  2022, \mn@doi [\mnras] {10.1093/mnras/stab3717}, \href
  {https://ui.adsabs.harvard.edu/abs/2022MNRAS.511..100B} {511, 100}

\bibitem[\protect\citeauthoryear{Brinchmann, Charlot, White, Tremonti,
  Kauffmann, Heckman  \& Brinkmann}{Brinchmann et~al.}{2004}]{B04}
Brinchmann J.,  Charlot S.,  White S. D.~M.,  Tremonti C.,  Kauffmann G.,
  Heckman T.,   Brinkmann J.,  2004, \mn@doi [\mnras]
  {10.1111/j.1365-2966.2004.07881.x}, 351, 1151

\bibitem[\protect\citeauthoryear{Chen et~al.,}{Chen et~al.}{2019}]{Chen19}
Chen Y.-M.,  et~al., 2019, \mn@doi [\mnras] {10.1093/mnras/stz2494}, 489, 5709

\bibitem[\protect\citeauthoryear{{{\'C}iprijanovi{\'c}}
  et~al.,}{{{\'C}iprijanovi{\'c}} et~al.}{2021}]{C21}
{{\'C}iprijanovi{\'c}} A.,  et~al., 2021, arXiv e-prints, \href
  {https://ui.adsabs.harvard.edu/abs/2021arXiv211100961C} {p. arXiv:2111.00961}

\bibitem[\protect\citeauthoryear{Conselice}{Conselice}{2003}]{Conselice03}
Conselice C.~J.,  2003, \mn@doi [\apjs] {10.1086/375001}, 147, 1

\bibitem[\protect\citeauthoryear{Conselice, Bershady  \& Jangren}{Conselice
  et~al.}{2000}]{Conselice00}
Conselice C.~J.,  Bershady M.~A.,   Jangren A.,  2000, \mn@doi [\apj]
  {10.1086/308300}, 529, 886

\bibitem[\protect\citeauthoryear{{Couch} \& {Sharples}}{{Couch} \&
  {Sharples}}{1987}]{CandS87}
{Couch} W.~J.,  {Sharples} R.~M.,  1987, \mn@doi [\mnras]
  {10.1093/mnras/229.3.423}, \href
  {https://ui.adsabs.harvard.edu/abs/1987MNRAS.229..423C} {229, 423}

\bibitem[\protect\citeauthoryear{{Davis}, {van de Voort}, {Rowlands},
  {McAlpine}, {Wild}  \& {Crain}}{{Davis} et~al.}{2019}]{Davis19}
{Davis} T.~A.,  {van de Voort} F.,  {Rowlands} K.,  {McAlpine} S.,  {Wild} V.,
   {Crain} R.~A.,  2019, \mn@doi [\mnras] {10.1093/mnras/stz180}, \href
  {https://ui.adsabs.harvard.edu/abs/2019MNRAS.484.2447D} {484, 2447}

\bibitem[\protect\citeauthoryear{{Dieleman}, {Willett}  \& {Dambre}}{{Dieleman}
  et~al.}{2015}]{Dieleman15}
{Dieleman} S.,  {Willett} K.~W.,   {Dambre} J.,  2015, \mn@doi [\mnras]
  {10.1093/mnras/stv632}, \href
  {https://ui.adsabs.harvard.edu/abs/2015MNRAS.450.1441D} {450, 1441}

\bibitem[\protect\citeauthoryear{{Dressler} \& {Gunn}}{{Dressler} \&
  {Gunn}}{1983}]{DG83}
{Dressler} A.,  {Gunn} J.~E.,  1983, \mn@doi [\apj] {10.1086/161093}, \href
  {https://ui.adsabs.harvard.edu/abs/1983ApJ...270....7D} {270, 7}

\bibitem[\protect\citeauthoryear{{Driver} et~al.,}{{Driver}
  et~al.}{2006}]{Driver06}
{Driver} S.~P.,  et~al., 2006, \mn@doi [\mnras]
  {10.1111/j.1365-2966.2006.10126.x}, \href
  {https://ui.adsabs.harvard.edu/abs/2006MNRAS.368..414D} {368, 414}

\bibitem[\protect\citeauthoryear{{Ellison}, {Viswanathan}, {Patton},
  {Bottrell}, {McConnachie}, {Gwyn}  \& {Cuillandre}}{{Ellison}
  et~al.}{2019}]{Ellison19}
{Ellison} S.~L.,  {Viswanathan} A.,  {Patton} D.~R.,  {Bottrell} C.,
  {McConnachie} A.~W.,  {Gwyn} S.,   {Cuillandre} J.-C.,  2019, \mn@doi
  [\mnras] {10.1093/mnras/stz1431}, \href
  {https://ui.adsabs.harvard.edu/abs/2019MNRAS.487.2491E} {487, 2491}

\bibitem[\protect\citeauthoryear{{Ellison}, {Wilkinson}, {Woo}, {Bickley},
  {Wild}, {Patton}, {Quai}  \& {Gwyn}}{{Ellison} et~al.}{2022}]{Ellison22}
{Ellison} S.~L.,  {Wilkinson} S.,  {Woo} J.,  {Bickley} R.~W.,  {Wild} V.,
  {Patton} D.~R.,  {Quai} S.,   {Gwyn} S.,  2022

\bibitem[\protect\citeauthoryear{{Ferreira}, {Conselice}, {Duncan}, {Cheng},
  {Griffiths}  \& {Whitney}}{{Ferreira} et~al.}{2020}]{Ferr20}
{Ferreira} L.,  {Conselice} C.~J.,  {Duncan} K.,  {Cheng} T.-Y.,  {Griffiths}
  A.,   {Whitney} A.,  2020, \mn@doi [\apj] {10.3847/1538-4357/ab8f9b}, \href
  {https://ui.adsabs.harvard.edu/abs/2020ApJ...895..115F} {895, 115}

\bibitem[\protect\citeauthoryear{{Ferreira}, {Conselice}, {Kuchner}  \&
  {Tohill}}{{Ferreira} et~al.}{2022}]{Ferreira22}
{Ferreira} L.,  {Conselice} C.~J.,  {Kuchner} U.,   {Tohill} C.-B.,  2022,
  \mn@doi [\apj] {10.3847/1538-4357/ac66ea}, \href
  {https://ui.adsabs.harvard.edu/abs/2022ApJ...931...34F} {931, 34}

\bibitem[\protect\citeauthoryear{{Goto}}{{Goto}}{2004}]{Goto04}
{Goto} T.,  2004, \mn@doi [A\&A] {10.1051/0004-6361:20041250}, 427, 125

\bibitem[\protect\citeauthoryear{Goto}{Goto}{2005}]{Goto05}
Goto T.,  2005, \mn@doi [\mnras] {10.1111/j.1365-2966.2005.08701.x}, 357, 937

\bibitem[\protect\citeauthoryear{{Goto}}{{Goto}}{2006}]{Goto06}
{Goto} T.,  2006, \mn@doi [\mnras] {10.1111/j.1365-2966.2006.10413.x}, \href
  {https://ui.adsabs.harvard.edu/abs/2006MNRAS.369.1765G} {369, 1765}

\bibitem[\protect\citeauthoryear{Goto}{Goto}{2007}]{Goto07}
Goto T.,  2007, \mn@doi [\mnras] {10.1111/j.1365-2966.2007.12227.x}, 381, 187

\bibitem[\protect\citeauthoryear{{Graham} \& {Driver}}{{Graham} \&
  {Driver}}{2005}]{Graham05}
{Graham} A.~W.,  {Driver} S.~P.,  2005, \mn@doi [PASA] {10.1071/AS05001}, \href
  {https://ui.adsabs.harvard.edu/abs/2005PASA...22..118G} {22, 118}

\bibitem[\protect\citeauthoryear{{Hani}, {Gosain}, {Ellison}, {Patton}  \&
  {Torrey}}{{Hani} et~al.}{2020}]{Hani2020}
{Hani} M.~H.,  {Gosain} H.,  {Ellison} S.~L.,  {Patton} D.~R.,   {Torrey} P.,
  2020, \mn@doi [\mnras] {10.1093/mnras/staa459}, \href
  {https://ui.adsabs.harvard.edu/abs/2020MNRAS.493.3716H} {493, 3716}

\bibitem[\protect\citeauthoryear{{Hogg}, {Masjedi}, {Berlind}, {Blanton},
  {Quintero}  \& {Brinkmann}}{{Hogg} et~al.}{2006}]{Hogg06}
{Hogg} D.~W.,  {Masjedi} M.,  {Berlind} A.~A.,  {Blanton} M.~R.,  {Quintero}
  A.~D.,   {Brinkmann} J.,  2006, \mn@doi [\apj] {10.1086/507172}, \href
  {https://ui.adsabs.harvard.edu/abs/2006ApJ...650..763H} {650, 763}

\bibitem[\protect\citeauthoryear{{Huertas-Company} et~al.,}{{Huertas-Company}
  et~al.}{2015}]{HC15}
{Huertas-Company} M.,  et~al., 2015, \mn@doi [\apjs]
  {10.1088/0067-0049/221/1/8}, \href
  {https://ui.adsabs.harvard.edu/abs/2015ApJS..221....8H} {221, 8}

\bibitem[\protect\citeauthoryear{Kauffmann et~al.,}{Kauffmann
  et~al.}{2003a}]{K03}
Kauffmann G.,  et~al., 2003a, \mn@doi [\mnras]
  {10.1046/j.1365-8711.2003.06291.x}, 341, 33

\bibitem[\protect\citeauthoryear{Kauffmann et~al.,}{Kauffmann
  et~al.}{2003b}]{K03a}
Kauffmann G.,  et~al., 2003b, \mn@doi [\mnras]
  {10.1111/j.1365-2966.2003.07154.x}, 346, 1055

\bibitem[\protect\citeauthoryear{{Lacey} \& {Cole}}{{Lacey} \&
  {Cole}}{1993}]{LC93}
{Lacey} C.,  {Cole} S.,  1993, \mn@doi [\mnras] {10.1093/mnras/262.3.627},
  \href {https://ui.adsabs.harvard.edu/abs/1993MNRAS.262..627L} {262, 627}

\bibitem[\protect\citeauthoryear{Lotz, Primack  \& Madau}{Lotz
  et~al.}{2004}]{Lotz04}
Lotz J.~M.,  Primack J.,   Madau P.,  2004, \mn@doi [\aj] {10.1086/421849},
  128, 163

\bibitem[\protect\citeauthoryear{Lotz, Jonsson, Cox  \& Primack}{Lotz
  et~al.}{2008}]{Lotz08}
Lotz J.~M.,  Jonsson P.,  Cox T.~J.,   Primack J.~R.,  2008, \mn@doi [\mnras]
  {10.1111/j.1365-2966.2008.14004.x}, 391, 1137

\bibitem[\protect\citeauthoryear{{Marinacci} et~al.,}{{Marinacci}
  et~al.}{2018}]{TNG4}
{Marinacci} F.,  et~al., 2018, \mn@doi [\mnras] {10.1093/mnras/sty2206}, \href
  {https://ui.adsabs.harvard.edu/abs/2018MNRAS.480.5113M} {480, 5113}

\bibitem[\protect\citeauthoryear{{Mendel}, {Simard}, {Palmer}, {Ellison}  \&
  {Patton}}{{Mendel} et~al.}{2014}]{Mendel14}
{Mendel} J.~T.,  {Simard} L.,  {Palmer} M.,  {Ellison} S.~L.,   {Patton} D.~R.,
   2014, \mn@doi [\apjs] {10.1088/0067-0049/210/1/3}, \href
  {https://ui.adsabs.harvard.edu/abs/2014ApJS..210....3M} {210, 3}

\bibitem[\protect\citeauthoryear{{Meusinger}, {Br{\"u}necke}, {Schalldach}  \&
  {in der Au}}{{Meusinger} et~al.}{2017}]{Meus2017}
{Meusinger} H.,  {Br{\"u}necke} J.,  {Schalldach} P.,   {in der Au} A.,  2017,
  \mn@doi [\aap] {10.1051/0004-6361/201629139}, \href
  {https://ui.adsabs.harvard.edu/abs/2017A&A...597A.134M} {597, A134}

\bibitem[\protect\citeauthoryear{{Mihos}}{{Mihos}}{1995}]{Mihos95}
{Mihos} J.~C.,  1995, \mn@doi [\apjl] {10.1086/187719}, \href
  {https://ui.adsabs.harvard.edu/abs/1995ApJ...438L..75M} {438, L75}

\bibitem[\protect\citeauthoryear{{Miller} \& {Owen}}{{Miller} \&
  {Owen}}{2001}]{Miller2001}
{Miller} N.~A.,  {Owen} F.~N.,  2001, \mn@doi [\apjl] {10.1086/320924}, \href
  {https://ui.adsabs.harvard.edu/abs/2001ApJ...554L..25M} {554, L25}

\bibitem[\protect\citeauthoryear{{Naiman} et~al.,}{{Naiman}
  et~al.}{2018}]{TNG5}
{Naiman} J.~P.,  et~al., 2018, \mn@doi [\mnras] {10.1093/mnras/sty618}, \href
  {https://ui.adsabs.harvard.edu/abs/2018MNRAS.477.1206N} {477, 1206}

\bibitem[\protect\citeauthoryear{{Nelson} et~al.,}{{Nelson}
  et~al.}{2018}]{TNG2}
{Nelson} D.,  et~al., 2018, \mn@doi [\mnras] {10.1093/mnras/stx3040}, \href
  {https://ui.adsabs.harvard.edu/abs/2018MNRAS.475..624N} {475, 624}

\bibitem[\protect\citeauthoryear{{Nielsen}, {Ridgway}, {De Propris}  \&
  {Goto}}{{Nielsen} et~al.}{2012}]{Nielsen12}
{Nielsen} D.~M.,  {Ridgway} S.~E.,  {De Propris} R.,   {Goto} T.,  2012,
  \mn@doi [\apjl] {10.1088/2041-8205/761/2/L16}, \href
  {https://ui.adsabs.harvard.edu/abs/2012ApJ...761L..16N} {761, L16}

\bibitem[\protect\citeauthoryear{{Osterbrock} \& {Ferland}}{{Osterbrock} \&
  {Ferland}}{2006}]{Osterbrock2006}
{Osterbrock} D.~E.,  {Ferland} G.~J.,  2006, {Astrophysics of gaseous nebulae
  and active galactic nuclei}

\bibitem[\protect\citeauthoryear{{Patton} et~al.,}{{Patton}
  et~al.}{2020}]{Patton2020}
{Patton} D.~R.,  et~al., 2020, \mn@doi [\mnras] {10.1093/mnras/staa913}, \href
  {https://ui.adsabs.harvard.edu/abs/2020MNRAS.494.4969P} {494, 4969}

\bibitem[\protect\citeauthoryear{Pawlik, Wild, Walcher, Johansson, Villforth,
  Rowlands, Mendez-Abreu  \& Hewlett}{Pawlik et~al.}{2016}]{Pawlik16}
Pawlik M.~M.,  Wild V.,  Walcher C.~J.,  Johansson P.~H.,  Villforth C.,
  Rowlands K.,  Mendez-Abreu J.,   Hewlett T.,  2016, \mn@doi [\mnras]
  {10.1093/mnras/stv2878}, 456, 3032

\bibitem[\protect\citeauthoryear{Pawlik et~al.,}{Pawlik
  et~al.}{2018}]{Pawlik18}
Pawlik M.~M.,  et~al., 2018, \mn@doi [\mnras] {10.1093/mnras/sty589}, 477, 1708

\bibitem[\protect\citeauthoryear{{Pawlik}, {McAlpine}, {Trayford}, {Wild},
  {Bower}, {Crain}, {Schaller}  \& {Schaye}}{{Pawlik} et~al.}{2019}]{Pawlik19}
{Pawlik} M.~M.,  {McAlpine} S.,  {Trayford} J.~W.,  {Wild} V.,  {Bower} R.,
  {Crain} R.~A.,  {Schaller} M.,   {Schaye} J.,  2019, \mn@doi [Nature
  Astronomy] {10.1038/s41550-019-0725-z}, \href
  {https://ui.adsabs.harvard.edu/abs/2019NatAs...3..440P} {3, 440}

\bibitem[\protect\citeauthoryear{{Pearson}, {Wang}, {Trayford}, {Petrillo}  \&
  {van der Tak}}{{Pearson} et~al.}{2019}]{Pearson19}
{Pearson} W.~J.,  {Wang} L.,  {Trayford} J.~W.,  {Petrillo} C.~E.,   {van der
  Tak} F.~F.~S.,  2019, \mn@doi [\aap] {10.1051/0004-6361/201935355}, \href
  {https://ui.adsabs.harvard.edu/abs/2019A&A...626A..49P} {626, A49}

\bibitem[\protect\citeauthoryear{{Pillepich} et~al.,}{{Pillepich}
  et~al.}{2018}]{TNG3}
{Pillepich} A.,  et~al., 2018, \mn@doi [\mnras] {10.1093/mnras/stx3112}, \href
  {https://ui.adsabs.harvard.edu/abs/2018MNRAS.475..648P} {475, 648}

\bibitem[\protect\citeauthoryear{{Poggianti} \& {Wu}}{{Poggianti} \&
  {Wu}}{2000}]{PogWu00}
{Poggianti} B.~M.,  {Wu} H.,  2000, \mn@doi [\apj] {10.1086/308243}, \href
  {https://ui.adsabs.harvard.edu/abs/2000ApJ...529..157P} {529, 157}

\bibitem[\protect\citeauthoryear{{Pracy}, {Kuntschner}, {Couch}, {Blake},
  {Bekki}  \& {Briggs}}{{Pracy} et~al.}{2009}]{Pracy09}
{Pracy} M.~B.,  {Kuntschner} H.,  {Couch} W.~J.,  {Blake} C.,  {Bekki} K.,
  {Briggs} F.,  2009, \mn@doi [\mnras] {10.1111/j.1365-2966.2009.14836.x},
  \href {https://ui.adsabs.harvard.edu/abs/2009MNRAS.396.1349P} {396, 1349}

\bibitem[\protect\citeauthoryear{{Pracy} et~al.,}{{Pracy}
  et~al.}{2013}]{Pracy2013}
{Pracy} M.~B.,  et~al., 2013, \mn@doi [\mnras] {10.1093/mnras/stt666}, \href
  {https://ui.adsabs.harvard.edu/abs/2013MNRAS.432.3131P} {432, 3131}

\bibitem[\protect\citeauthoryear{{Quintero} et~al.,}{{Quintero}
  et~al.}{2004}]{Quintero2004}
{Quintero} A.~D.,  et~al., 2004, \mn@doi [\apj] {10.1086/380601}, \href
  {https://ui.adsabs.harvard.edu/abs/2004ApJ...602..190Q} {602, 190}

\bibitem[\protect\citeauthoryear{{Rodriguez-Gomez} et~al.,}{{Rodriguez-Gomez}
  et~al.}{2015}]{RG15}
{Rodriguez-Gomez} V.,  et~al., 2015, \mn@doi [\mnras] {10.1093/mnras/stv264},
  \href {https://ui.adsabs.harvard.edu/abs/2015MNRAS.449...49R} {449, 49}

\bibitem[\protect\citeauthoryear{{Rodriguez-Gomez} et~al.,}{{Rodriguez-Gomez}
  et~al.}{2019}]{RG19}
{Rodriguez-Gomez} V.,  et~al., 2019, \mn@doi [\mnras] {10.1093/mnras/sty3345},
  \href {https://ui.adsabs.harvard.edu/abs/2019MNRAS.483.4140R} {483, 4140}

\bibitem[\protect\citeauthoryear{Rowlands, Wild, Nesvadba, Sibthorpe, Mortier,
  Lehnert  \& da Cunha}{Rowlands et~al.}{2015}]{Rowlands15}
Rowlands K.,  Wild V.,  Nesvadba N.,  Sibthorpe B.,  Mortier A.,  Lehnert M.,
  da Cunha E.,  2015, \mn@doi [\mnras] {10.1093/mnras/stu2714}, 448, 258

\bibitem[\protect\citeauthoryear{{Rowlands} et~al.,}{{Rowlands}
  et~al.}{2018}]{Rowlands18}
{Rowlands} K.,  et~al., 2018, \mn@doi [\mnras] {10.1093/mnras/sty1916}, \href
  {https://ui.adsabs.harvard.edu/abs/2018MNRAS.480.2544R} {480, 2544}

\bibitem[\protect\citeauthoryear{{Salim} et~al.,}{{Salim}
  et~al.}{2007}]{Salim07}
{Salim} S.,  et~al., 2007, \mn@doi [\apjs] {10.1086/519218}, \href
  {https://ui.adsabs.harvard.edu/abs/2007ApJS..173..267S} {173, 267}

\bibitem[\protect\citeauthoryear{{Sazonova} et~al.,}{{Sazonova}
  et~al.}{2021}]{Saz21}
{Sazonova} E.,  et~al., 2021, \mn@doi [\apj] {10.3847/1538-4357/ac0f7f}, \href
  {https://ui.adsabs.harvard.edu/abs/2021ApJ...919..134S} {919, 134}

\bibitem[\protect\citeauthoryear{{S{\'e}rsic}}{{S{\'e}rsic}}{1963}]{Sersic63}
{S{\'e}rsic} J.~L.,  1963, Boletin de la Asociacion Argentina de Astronomia La
  Plata Argentina, \href
  {https://ui.adsabs.harvard.edu/abs/1963BAAA....6...41S} {6, 41}

\bibitem[\protect\citeauthoryear{{Simard}, {Mendel}, {Patton}, {Ellison}  \&
  {McConnachie}}{{Simard} et~al.}{2011}]{Simard11}
{Simard} L.,  {Mendel} J.~T.,  {Patton} D.~R.,  {Ellison} S.~L.,
  {McConnachie} A.~W.,  2011, \mn@doi [\apjs] {10.1088/0067-0049/196/1/11},
  \href {https://ui.adsabs.harvard.edu/abs/2011ApJS..196...11S} {196, 11}

\bibitem[\protect\citeauthoryear{{Smail}, {Morrison}, {Gray}, {Owen}, {Ivison},
  {Kneib}  \& {Ellis}}{{Smail} et~al.}{1999}]{Smail99}
{Smail} I.,  {Morrison} G.,  {Gray} M.~E.,  {Owen} F.~N.,  {Ivison} R.~J.,
  {Kneib} J.~P.,   {Ellis} R.~S.,  1999, \mn@doi [\apj] {10.1086/307934}, \href
  {https://ui.adsabs.harvard.edu/abs/1999ApJ...525..609S} {525, 609}

\bibitem[\protect\citeauthoryear{{Snyder}, {Cox}, {Hayward}, {Hernquist}  \&
  {Jonsson}}{{Snyder} et~al.}{2011}]{Snyder2011}
{Snyder} G.~F.,  {Cox} T.~J.,  {Hayward} C.~C.,  {Hernquist} L.,   {Jonsson}
  P.,  2011, \mn@doi [\apj] {10.1088/0004-637X/741/2/77}, \href
  {https://ui.adsabs.harvard.edu/abs/2011ApJ...741...77S} {741, 77}

\bibitem[\protect\citeauthoryear{{Somerville} \& {Dav{\'e}}}{{Somerville} \&
  {Dav{\'e}}}{2015}]{SomDave15}
{Somerville} R.~S.,  {Dav{\'e}} R.,  2015, \mn@doi [\araa]
  {10.1146/annurev-astro-082812-140951}, \href
  {https://ui.adsabs.harvard.edu/abs/2015ARA&A..53...51S} {53, 51}

\bibitem[\protect\citeauthoryear{Springel et~al.,}{Springel
  et~al.}{2017}]{TNG1}
Springel V.,  et~al., 2017, \mn@doi [\mnras] {10.1093/mnras/stx3304}, 475, 676

\bibitem[\protect\citeauthoryear{Strateva et~al.,}{Strateva
  et~al.}{2001}]{Strateva_2001}
Strateva I.,  et~al., 2001, \mn@doi [\aj] {10.1086/323301}, 122, 1861

\bibitem[\protect\citeauthoryear{{Tremonti}, {Moustakas}  \&
  {Diamond-Stanic}}{{Tremonti} et~al.}{2007}]{Tremonti07}
{Tremonti} C.~A.,  {Moustakas} J.,   {Diamond-Stanic} A.~M.,  2007, \mn@doi
  [\apjl] {10.1086/520083}, \href
  {https://ui.adsabs.harvard.edu/abs/2007ApJ...663L..77T} {663, L77}

\bibitem[\protect\citeauthoryear{{Werle} et~al.,}{{Werle}
  et~al.}{2022}]{Werle22}
{Werle} A.,  et~al., 2022, \mn@doi [\apj] {10.3847/1538-4357/ac5f06}, \href
  {https://ui.adsabs.harvard.edu/abs/2022ApJ...930...43W} {930, 43}

\bibitem[\protect\citeauthoryear{Wild, Kauffmann, Heckman, Charlot, Lemson,
  Brinchmann, Reichard  \& Pasquali}{Wild et~al.}{2007}]{Wild07}
Wild V.,  Kauffmann G.,  Heckman T.,  Charlot S.,  Lemson G.,  Brinchmann J.,
  Reichard T.,   Pasquali A.,  2007, \mn@doi [\mnras]
  {10.1111/j.1365-2966.2007.12256.x}, 381, 543

\bibitem[\protect\citeauthoryear{Wild, Walcher, Johansson, Tresse, Charlot,
  Pollo, Le~Fèvre  \& De~Ravel}{Wild et~al.}{2009}]{Wild09}
Wild V.,  Walcher C.~J.,  Johansson P.~H.,  Tresse L.,  Charlot S.,  Pollo A.,
  Le~Fèvre O.,   De~Ravel L.,  2009, \mn@doi [\mnras]
  {10.1111/j.1365-2966.2009.14537.x}, 395, 144

\bibitem[\protect\citeauthoryear{Wild, Heckman  \& Charlot}{Wild
  et~al.}{2010}]{Wild10}
Wild V.,  Heckman T.,   Charlot S.,  2010, \mn@doi [\mnras]
  {10.1111/j.1365-2966.2010.16536.x}, 405, 933

\bibitem[\protect\citeauthoryear{{Wild} et~al.,}{{Wild} et~al.}{2020}]{Wild20}
{Wild} V.,  et~al., 2020, \mn@doi [\mnras] {10.1093/mnras/staa674}, \href
  {https://ui.adsabs.harvard.edu/abs/2020MNRAS.494..529W} {494, 529}

\bibitem[\protect\citeauthoryear{Wuyts et~al.,}{Wuyts et~al.}{2011}]{Wuyts2011}
Wuyts S.,  et~al., 2011, \mn@doi [\apj] {10.1088/0004-637x/742/2/96}, 742, 96

\bibitem[\protect\citeauthoryear{{Yan} et~al.,}{{Yan} et~al.}{2009}]{Yan09}
{Yan} R.,  et~al., 2009, \mn@doi [\mnras] {10.1111/j.1365-2966.2009.15192.x},
  \href {https://ui.adsabs.harvard.edu/abs/2009MNRAS.398..735Y} {398, 735}

\bibitem[\protect\citeauthoryear{{Yang}, {Zabludoff}, {Zaritsky}  \&
  {Mihos}}{{Yang} et~al.}{2008}]{Yang08}
{Yang} Y.,  {Zabludoff} A.~I.,  {Zaritsky} D.,   {Mihos} J.~C.,  2008, \mn@doi
  [\apj] {10.1086/591656}, \href
  {https://ui.adsabs.harvard.edu/abs/2008ApJ...688..945Y} {688, 945}

\bibitem[\protect\citeauthoryear{Yesuf, Faber, Trump, Koo, Fang, Liu, Wild  \&
  Hayward}{Yesuf et~al.}{2014}]{Yesuf_2014}
Yesuf H.~M.,  Faber S.~M.,  Trump J.~R.,  Koo D.~C.,  Fang J.~J.,  Liu F.~S.,
  Wild V.,   Hayward C.~C.,  2014, \mn@doi [\apj] {10.1088/0004-637x/792/2/84},
  792, 84

\bibitem[\protect\citeauthoryear{{Zabludoff}, {Zaritsky}, {Lin}, {Tucker},
  {Hashimoto}, {Shectman}, {Oemler}  \& {Kirshner}}{{Zabludoff}
  et~al.}{1996}]{Zab96}
{Zabludoff} A.~I.,  {Zaritsky} D.,  {Lin} H.,  {Tucker} D.,  {Hashimoto} Y.,
  {Shectman} S.~A.,  {Oemler} A.,   {Kirshner} R.~P.,  1996, \mn@doi [\apj]
  {10.1086/177495}, \href
  {https://ui.adsabs.harvard.edu/abs/1996ApJ...466..104Z} {466, 104}

\bibitem[\protect\citeauthoryear{{Zheng}, {Wild}, {Lah{\'e}n}, {Johansson},
  {Law}, {Weaver}  \& {Jimenez}}{{Zheng} et~al.}{2020}]{Zheng2020}
{Zheng} Y.,  {Wild} V.,  {Lah{\'e}n} N.,  {Johansson} P.~H.,  {Law} D.,
  {Weaver} J.~R.,   {Jimenez} N.,  2020, \mn@doi [\mnras]
  {10.1093/mnras/staa2358}, \href
  {https://ui.adsabs.harvard.edu/abs/2020MNRAS.498.1259Z} {498, 1259}

\makeatother
\end{thebibliography}

\appendix
\newpage
\section{The Effect of Recent Central Starbursts on Asymmetry}
    
    \label{central}
    
    In Section \ref{Results}, we show that shape asymmetry, which gives additional weight to low-surface brightness features, detects significant fractions of disturbed morphologies in each of the PSB populations. However, we also found that asymmetry, which is commonly used to identify mergers \citep{Conselice03}, does not find significant fractions of disturbed morphologies in the PSB samples. In fact, the asymmetry of both PSB samples were suppressed relative to their star-forming controls which are expected to have some asymmetric features like spiral arms, clumpy star formation, and dust obstruction. We argue this is a natural result of our PSB definition; our PSBs are selected on the basis of spectra from fibres placed on the central region of each galaxy\footnote{To use the terms from \citet{Chen19}, these PSBs would be considered central PSBs (CPSBs) as opposed to ring PSBs (RPSBs) or irregular PSBs (IPSBs) which can only be identified with spatially resolved spectroscopy.}. Since our PSB selection uses information only from the central region, these galaxies have had, by definition, a recent burst of central star-formation. A central burst in star formation would create a bright, azimuthally symmetric core which would systematically drive down the asymmetry values which are weighted by each pixel's intensity. To confirm this idea, we explore the concentration and Sérsic index of our PSB samples in relation to their star-forming controls.
    
    The concentration, C, of a galaxy is quantified by \citet{Conselice03} as:
    
    \begin{equation}
        C = 5\text{log}\left( \frac{r_{80}}{r_{20}}\right),
    \end{equation}
    
    \noindent where $r_{20}$ and $r_{80}$ are the radii of circular apertures containing 20 and 80 per cent of the galaxy’s light, respectively. Elliptical galaxies tend to be more concentrated ($C = 4.4\pm0.3$) than spiral galaxies \citep[$C = 3.9\pm0.5$;][]{Conselice03}. 
    
    \begin{figure}
        \centering
        \includegraphics[width=1\linewidth]{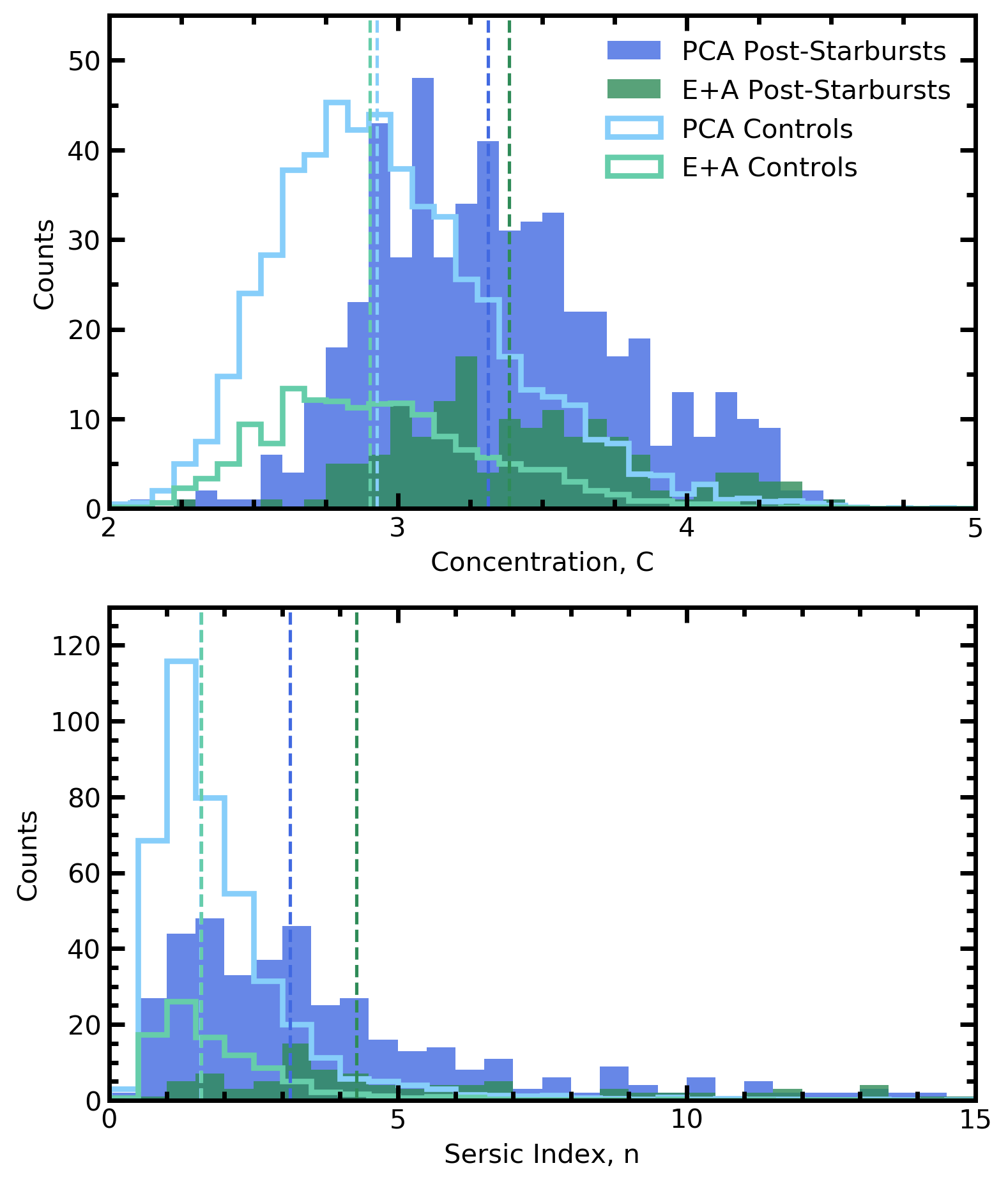}
        \caption{The distributions of concentration (top) and Sérsic index (bottom) for the PCA PSBs (blue) and their controls (unfilled light-blue) and for the E+A PSBs (green) and their controls (unfilled light green). Note that since there are ten controls for each post-starburst, the histograms of the controls are weighted by a factor of 1/10 so they can be compared directly to the post-starburst distributions and that PSBs and their controls in the lower panel are required to have un-flagged Sérsic fits from \texttt{statmorph}.}
        \label{C-n-dists}
    \end{figure}
    
    Figure \ref{C-n-dists} shows that both samples of PSBs have much higher concentrations than their star-forming controls. However, this could be a result of PSBs transitioning elliptical to morphologies, which tend to have both higher concentrations and lower asymmetries, or caused by a recent central starburst increasing the concentration of the stellar light profile.
    
    We can break the degeneracy between elliptical-type morphologies and very bright nuclei using the Sérsic index. A 1-dimensional Sérsic profile \citep{Sersic63} can be used to approximate the intensity of light (as a function of radius) of a galaxy:
    
    \begin{equation}
    \label{Sersic}
        I(R) = I_e \exp\left\{-b_n\left[\left( \frac{R}{R_e}\right)^{-1/n} - 1\right]\right\}
    \end{equation}
    
    \noindent where $R_e$ is the radius within which half of the galaxy's light is contained, $I_e$ is the intensity of the light at the radius, $R_e$, $n$ is the Sérsic index and $b_n$ is a coefficient determined by the Sérsic index \citep[see][]{Graham05}.
    
    The Sérsic index, $n$, describes the rate at which light decreases from its central peak. By this definition a galaxy with a Sérsic index of $n = 1$ has light that drops off as an exponential function and as $n$ increases, the curvature of the light profile tends towards linearity. The curvature of the profile, controlled by the Sérsic index, $n$, is strongly correlated to the morphological classification of the galaxy; galaxies with Sérsic index $n \thicksim 1$ tend to be spiral galaxies and galaxies with $2 \lesssim n \lesssim 4$ tend to be elliptical \citep{Sersic63}.
    
    Figure \ref{C-n-dists} shows that the PCA PSBs exhibit a wide range of Sérsic indices, indicating both disks and ellipticals. In fact, both PSB samples have bimodal Sérsic index distributions, with distinct peaks at $n\thicksim$1 and $n\thicksim$4. Compare this with the clear unimodal distribution around $n\thicksim$1 of their star-forming controls and it is clear that post-starburst galaxies exhibit a wide range of morphological types, not just elliptical. This rules out elliptical morphologies dominating the PSB samples and driving the observed increase in concentration and decrease in asymmetry.
    
    Lastly, more than half of the PSBs in both samples have Sérsic indices much higher than a typical elliptical galaxy ($n \gtrsim 4$). These high Sérsic indices are a result of trying to fit galaxies with a bright centralized core with a single Sérsic profile. The results presented here are consistent with the results of \citet{Goto05} who found that PSBs tend to have very bright nuclei based on visual inspection and with the findings of \citet{Wuyts2011} who find higher Sérsic indices for starburst galaxies with enhanced SFR.

\section{A Morphological Comparison Between CFIS and SDSS}
\label{appA}

\begin{figure}
        \centering
        \includegraphics[width=\linewidth]{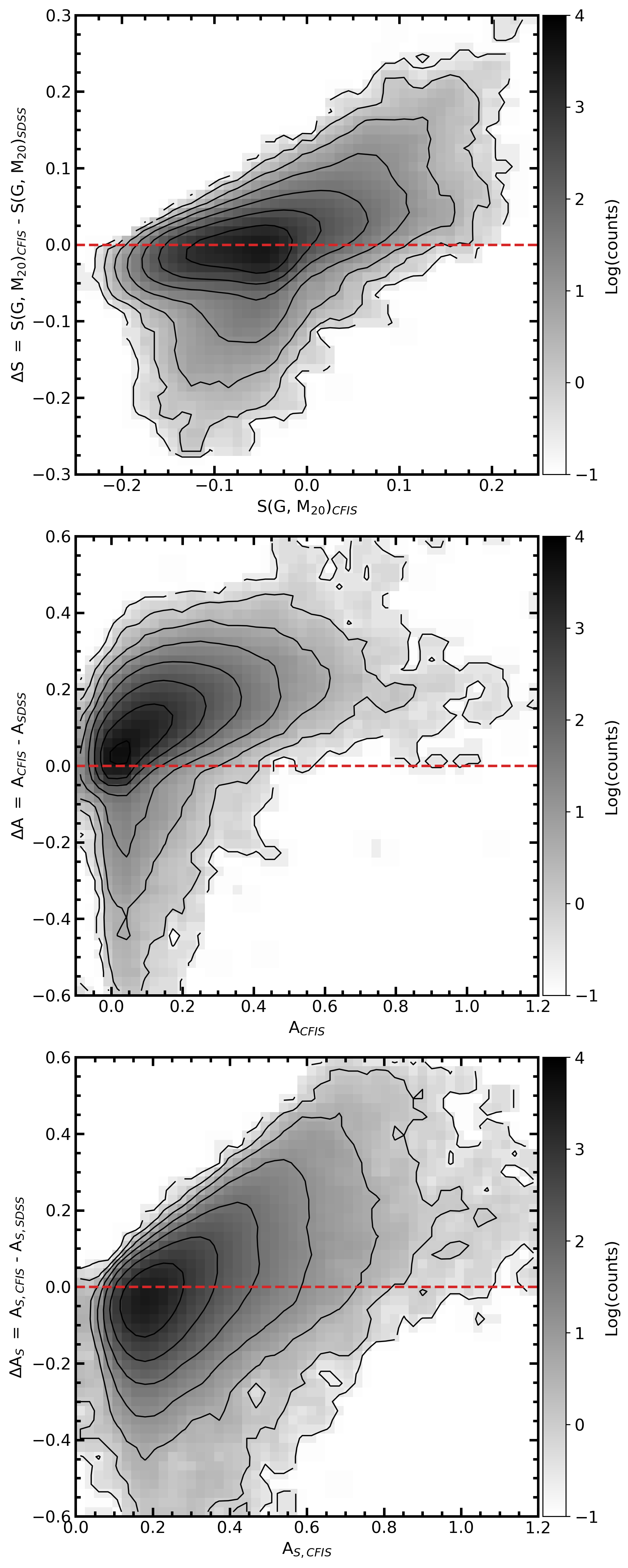}
        \caption{The differences between non-parametric morphologies as computed on a galaxy-by-galaxy basis using CFIS and SDSS \emph{r}-band imaging. The dashed red red line denotes the line of equality between the two surveys and the colour gradient of the histogram denotes the number of galaxies in each bin in log scale. The histogram and contours are smoothed by a 3$\times$3 kernel.}
        \label{CvS_summary}
    \end{figure}

    We have computed morphology statistics with \texttt{statmorph} for both the SDSS and CFIS \emph{r}-band imaging for the $\thicksim$168,000 galaxies in the overlap between the two surveys (see Sections \ref{methods} and \ref{mfrac_cvs}). In Figure \ref{CvS_summary}, we present the difference between the non-parametric morphology statistics used in this work as derived using CFIS and SDSS imaging.
    
     Looking first at the difference between S(G, M$_{20}$) in CFIS and SDSS shown in the top panel of Figure \ref{CvS_summary}, the vast majority ($\thicksim$95\%) of galaxies differ between the two surveys by less than 0.05 and the scatter is symmetric about 0. By visually inspecting the images of the small number of cases ($<1\%$) in the top right corner of this panel where S(G, M$_{20}$) $> 0$ in CFIS and the difference between the two surveys is large ($\Delta S > 0.1$), we find they are caused by foreground stars and close pairs that are not deblended from the target galaxy in CFIS because of an extended diffuse stellar halo but are deblended in SDSS, leading to a lower value of S(G, M$_{20}$). Galaxies found in the bottom left of this plot have the inverse deblending issue where objects are not deblended in SDSS because of PSF blurring. 
     
     The middle panel of Figure \ref{CvS_summary} shows that most galaxies have low values of asymmetry in both surveys, but those with high asymmetry in CFIS are higher in CFIS than SDSS. The small number of galaxies ($<1\%$) with $\Delta A < -0.1$ are caused by close pairs and foreground stars that are successfully deblended from the galaxy in CFIS but not in SDSS. 
     
     Finally, we explore the difference in shape asymmetry between the two surveys. At high values of $A_S$ in CFIS, the values are consistently higher in CFIS than in SDSS since low surface brightness azimuthally asymmetric features are more likely to be detected in the deeper survey. However, there are a significant number of galaxies with higher shape asymmetries in SDSS than in CFIS which are often caused by deblending errors but many of which are bona fide mergers. Like the example galaxy in Section \ref{IQIll}, the shape asymmetry will actually be lower when using deeper imaging if there is a diffuse stellar halo that is more symmetric than the slightly brighter tidal tails. In such a case, the binary mask in SDSS includes only the asymmetric tidal tails, while in CFIS, the binary mask includes the faint and symmetric diffuse stellar halo. Another peculiar difference in the shape asymmetry measurements between the surveys is that inherently symmetric galaxies are assigned a higher shape asymmetry in SDSS than in CFIS. This is caused by the lower resolution of the imaging creating a more pixelated binary mask which has a higher probability of having asymmetric pixels. 

\bsp	
\label{lastpage}
\end{document}